%% file: ms-emapj.tex
\documentclass[iop]{emulateapj}

\usepackage{verbatim}
\usepackage{natbib}
\newcommand \degree {^{\circ} }

\newcommand \hii {\ion{H}{2} }
\newcommand \micronm {\,\mu{\rm m} } 
\newcommand \jmag {\,$J$}
\newcommand \hmag {\,$H$} 
\newcommand \kmag {\,$K_{S}$} 
\newcommand \jh {\,$J-H$} 
\newcommand \hk {\,$H-K_{S}$}

\begin{document}
\title{OB Associations at the Upper End of the Milky Way Luminosity
  Function} \author{Mubdi Rahman\altaffilmark{1,2}, Christopher
  D. Matzner\altaffilmark{1} \& Dae-Sik Moon\altaffilmark{1,3}}
\email{mubdi@pha.jhu.edu} \altaffiltext{1}{Department of Astronomy \&
  Astrophysics, University of Toronto, 50 St.~George Street, Toronto,
  Ontario, M5S 3H8, Canada} \altaffiltext{2}{Department of Physics and
  Astronomy, Johns Hopkins University, Baltimore, MD 21218, USA}
\altaffiltext{3}{Brain Pool Scholar, Korea Astronomy and Space Science
  Institute, Daejeon, Korea, 305-348}

\begin{abstract}

  The Milky Way's most luminous, young and massive (M $\ga 10^4$
  M$_{\sun}$) star clusters and OB associations have largely evaded
  detection despite knowledge of their surrounding \hii regions. We
  search for these clusters and associations within the 40 star
  forming complexes from Rahman \& Murray in the 13 most luminous WMAP
  free-free emission sources of the Galaxy. Selecting for objects with
  the dust-reddened colors of OB stars, we identify new candidate
  associations using the 2MASS point source catalog. In 40 star
  forming complexes searched, 22 contain cluster/association
  candidates with sizes and masses in the range of
  3\arcmin--26\arcmin\, and $10^{2.3}$--$10^{5}$ M$_{\sun}$.  Of the
  22 candidates, at least 7 have estimated masses $\ga 10^4$
  M$_{\sun}$, doubling the number of such massive clusters known in
  the Galaxy. Applying our method to a statistically similar set of
  test locations, we estimate that $3.0 \pm 0.6$ of our 22 candidate
  associations are unrelated to the star forming complexes. In
  addition, the apparent extinctions of our candidate associations
  correlate well with the predictions from a Galactic model. These
  facts, along with the clear detection of a known OB association and
  the previous spectral verification of one cluster found by this
  method, validate our method.  Only one of the searched WMAP sources
  remains without a candidate. In 8 of the most luminous WMAP sources,
  the candidate associations can account for the observed free-free
  flux. With our new compilation, the Galactic census of young,
  massive stellar associations may now be about half complete.

\end{abstract}


\section{Introduction}

A spiral galaxy's metabolism is dominated by the effects of its most
massive (M $\ga 10^4$ M$_{\sun}$) young stellar clusters and OB
associations. Not only are these clusters and OB associations the
birthplaces of a substantial fraction of a galaxy's stars, but they
also dominate the energetic feedback within the galactic disk and
drive disk-halo interactions through the production of superbubbles
\citep{mckee97, chu08, weaver77, norman89, reynolds01}. While the
upper end of the cluster mass function (CMF) has been derived through
Galactic \hii region observations \citep{mckee97, murray10}, the
powering clusters of the \hii regions have mostly evaded
detection. This is primarily due to the large distance and
extinguishing column through the Galactic disk. Although a number of
members of the upper end of the CMF have been identified and
characterized, including NGC 3603 \citep{stolte04} and the Galactic
Center clusters \citep{figer99}, the census of the upper end is far
from complete. This poses a problem for any global studies of the
Galaxy's energy budget, star formation properties, turbulence, and
overall ecology. Additionally, more complete knowledge of the upper
end of the CMF is required for detailed comparison to other galaxies,
critical to further understanding the role of star formation in galaxy
evolution \citep{whitmore07, larsen09}.

Recent work by \citet[][hereafter RM10]{rahman10} with data from the
\emph{Wilkinson Microwave Anisotropy Probe} (WMAP) and the
\emph{Spitzer} GLIMPSE survey has identified 40 star forming complexes
(SFCs) within the Galaxy's most luminous free-free emission
sources. Only one of the SFCs has a previously identified powering
cluster (NGC 3603), indicating that the complexes may be harboring
unknown luminous young massive clusters and OB associations. Notably,
\citet{rahman11a} identified a candidate association for the most
luminous SFC (the Dragonfish Nebula) based on the density of
near-infrared (NIR) point sources, filtering non-member field stars
using an extinction-based color-cut. The candidate association is
located 9.7 kpc through the Galactic disk with a line-of-sight
extinction $1.0 <$ A$_{K} < 1.4$ mag. Without the NIR color selection
process, the association would remain hidden to visual or automated
cluster search methods. \citet{rahman11b} confirmed this association
with NIR spectra of a sample of its brightest members.

In this paper, we expand upon the method of \citet{rahman11a} to
search for young massive clusters or OB associations in all of the
SFCs identified in RM10. In \S\ref{sect:prev}, we discuss previous
cluster searches.  In \S\ref{sect:sfcs}, we outline the nature of the
target SFCs.  In \S\ref{sect:data}, we discuss the use of the Two
Micron All-Sky Survey (2MASS) for cluster searches and present the
cluster search technique. In \S\ref{sect:candclust}, we present our
new candidate clusters and associations, discuss their statistical
significance, and test the method through which they are
identified. In \S\ref{sect:phys}, we estimate masses and luminosities
for the candidates. In \S\ref{sect:gcmf}, we produce and discuss the
most recent census of Galactic young massive clusters and OB
associations. We summarize our conclusions in
\S\ref{sect:conclusions}.

\section{Previous Cluster Searches}
\label{sect:prev}

Star clusters have historically been discovered in the visible
wavelengths. However, these discoveries have been limited to the
closest cluster populations due to the substantial dust obscuration
through the disk. The advent of large NIR surveys has enabled the
search for star clusters and OB associations despite this
obscuration. Specifically, the 2MASS (1.2--2.2 $\micronm$) and the
Spitzer Galactic Legacy Infrared Mid-Plane Survey Extraordinaire
(GLIMPSE, 3.6--8.0 $\micronm$) have produced catalogs covering large
areas of the Galactic plane in wavelengths where stellar flux
dominates but extinction effects are minimized, although not
completely eliminated \citep{skrutskie06, benjamin03}.

\citet{dutra03} and \citet{bica03} have systematically searched for
star clusters in 2MASS, focused around optical and radio
nebulae--primarily \hii regions. Their search used visual
identification of clustering on small images (5\arcmin -- 15\arcmin{}
square) centered around the nebulae. Though this process, 346 new
cluster candidates were identified. However, these efforts only
searched for clusters around known nebulae, missing those not
coincident with nebular emission (see \S \ref{sect:sfcs}). In a more
automated manner, \citet{froebrich07} used the total on-sky stellar
density of 2MASS point sources within $\vert b \vert < 20 \degr $ to
identify 1788 cluster candidates, of which 1021 are new. This method
is less biased towards nebula-coincident clusters, but more
susceptible to false positives. They estimated that their
contamination rate of chance overdensities within their catalog is
approximately 50\%. In both these cases, the searches are most
sensitive to dense clusters, rather than diffuse regions. For
instance, \citet{froebrich07} only detected a small clustered region
within Cygnus OB2, rather than the entire OB association with a
half-light radius of 13\arcmin, which is readily identifiable with
2MASS \citep{knod00}. Over 96\% of the \citet{froebrich07} catalog
candidates have a total central cluster density of greater than four
sources arcminute$^{-2}$, and over half greater than 10 sources
arcminute$^{-2}$. For the highly veiled case where only the most
luminous stars are visible, typical source densities of $10^{4}$
M$_{\sun}$ OB associations are expected to be $\sim 1.0 (R/10 \textrm{
  pc})^{-2} (D/ 10\textrm{ kpc})^{0.84}$ sources arcminute$^{-2}$,
where $R$ and $D$ are the association radius and distance
\citep{rahman11a}.  For comparison, the typical stellar density of a
2MASS field in the Galactic Plane is between 14 and 17 sources
arcminute$^{-2}$ (\S\ref{subsect:2mass}).  Consequently, these
previous 2MASS searches are insensitive to more dispersed OB
associations. Similarly, \citet{mercer05} identified 92 new cluster
candidates using similar automated and visual identification methods
with the GLIMPSE point source catalog. All of the GLIMPSE cluster
candidates have diameters of 3\arcmin{} or less, indicating the same
bias towards dense clusters. Since there are likely many more
associations with diameters $>3$\arcmin{}, such as Cygnus OB2, a
systematic search optimized towards larger, more massive OB
associations is required to find them.

For the candidate clusters identified in the works of \citet{dutra03},
\citet{bica03}, and \citet{mercer05}, further photometric and
spectroscopic studies have found high rates of false positive clusters
\citep{borissova05, ivanov05}. Those clusters that are confirmed tend
to have masses in the range of 10--1,000 M$_{\sun}$ and contain no
stars with spectral types earlier than B, so cannot be significant
ionization sources \citep{hanson08, soares08}. This suggests that most
of the galaxy's largest ionizing clusters have not been identified.

The recent work by \citet{bica11} indicates that when candidate
clusters are poorly populated or projected against dense stellar
backgrounds, discerning a real cluster from a field density
fluctuation (or asterism) is difficult or impossible. The effect poses
a problem in the search for the most luminous clusters in the Galaxy
located at significant distances; only the brightest members would be
visible due to the substantial extinction and distances through the
Galactic disk, even in the infrared regime. This creates the
appearance of a poorly populated cluster. Further, most, if not all,
of these clusters and associations should be located inside the
Galactic disk, leading to a substantial confusing stellar background
population. Therefore, these previous searches are best suited to
finding nearby, compact clusters far from the Galactic mid-plane.

In summary, previous automated cluster searches are sensitive to
compact clusters, like NGC 3603, but not to more diffuse associations
like Cygnus OB2 and the Dragonfish. To thoroughly catalog the Galaxy's
most luminous clusters using NIR surveys, one must exclude a large
fraction of the field giants.

\section{WMAP-identified Star Forming Complexes}
\label{sect:sfcs}

To identify the locations of the most luminous OB associations, we use
the identification of their parental SFCs from the WMAP and Spitzer
GLIMPSE survey \citep[][ RM10]{murray10}. The WMAP mission provides
all-sky, high sensitivity ($\sim \mu$K) microwave maps with five bands
in the $23.5 - 90$ GHz range with a relatively large ($\sim 1\degr$)
beam size. The five bands permit the separation of each of the
Galactic ``foreground'' emission components; synchrotron emission,
thermal dust emission, and free-free emission. The free-free emission
almost entirely arises from reprocessed ionizing photons produced by
the young, luminous (and massive) stars that are found {\it en masse}
in the most luminous young clusters and OB
associations. \citet{murray10} identify free-free emission sources
through the WMAP foreground map, finding that 14 sources are
responsible for one-third of the total galactic ionizing
luminosity. These regions are expected to be powered by young, very
massive clusters or OB associations that contain sufficient O-type
stars to produce the observed free-free emission. Within these 14
sources, only two have had previously identified powering clusters and
associations, specifically the Galactic Center region and NGC 3603
\citep{murray10}. Also notable is that many of the previously known
massive OB associations, such as Carina and Cygnus OB2, are not
enclosed by these 13 most luminous regions. This suggests the
existence of ionizing clusters or associations more luminous than
those previously identified.

The large ($\ga 1 \degr$) radii of the WMAP sources leads to
significant confusion along the line of sight, especially in the
sources towards the inner Galaxy. RM10 use the Spitzer GLIMPSE
$8\micronm$ emission maps and radio recombination line radial
velocities to separate the large WMAP sources into SFCs for the 13
most luminous sources from \citet{murray10}, excluding the Galactic
Center region. The observed SFCs differ in structure from the
traditional model of a Str\"{o}mgren sphere \hii region: the complexes
often appear as limb-brightened shells with the strongest nebular
emission tracing large shells, typically $5\arcmin$ to $30 \arcmin$ in
radius, rather than in the central region. This gives the impression
that a powerful central association has evacuated a zone of neutral
gas, which it illuminates and ionizes. The brightest knots of emission
in these shells, which often coincide with previously identified
\ion{H}{2} regions, could represent sites of triggered massive star
formation. Further, \citet{murray10} find that the majority of
ionizing photons escape from the shells and produce diffuse ionization
surrounding the entire region, often with radii $> 1\degr$, an effect
previously inferred in the literature \citep{anan85, mckee97}. While
the classical Str\"{o}mgren sphere interpretation of the SFC structure
would suggest that the powering source should be located towards the
center of the bright emission, this scenario suggests that the cluster
is not expected to be coincident with the nebular emission for any but
the youngest and smallest complexes.

Within the 13 investigated WMAP sources, RM10 identify 40 discrete
SFCs with distances ranging from 2.5 to 15 kpc and semimajor axes
between $2\arcmin$ and $67\arcmin$. In some cases, the kinematic
distance ambiguity is not resolved and two possible distances are
identified. Since the WMAP sources are distant and projected against
the crowded Galactic plane, we expect a great deal of confusion along
the line of sight: while the inferred luminosities of these sources
suggest that they contain massive clusters, they also likely contain a
number of less-massive sources along the line of sight. Due to the
large fraction of ionizing photons escaping the central area of the
complex to radii similar to the distances between complexes, we cannot
disentangle the different flux components from each of the
sources. This makes it difficult {\it a priori} to determine which of
the complexes are home to the massive clusters using ionization-based
measures. In this work, we search for massive clusters and
associations in all of the identified SFCs.

Previously, we have directed our attention to the most luminous of the
SFCs, the Dragonfish Nebula. We developed a method using the 2MASS
point source catalog to identify a candidate for the Galaxy's most
luminous OB association, with a total stellar mass of 10$^{5}$
M$_{\sun}$ \citep{rahman11a}. Follow-up NIR spectroscopy of bright
stars within the candidate association confirmed the existence of
Dragonfish Association with membership statistics consistent with the
candidate's expected value \citep{rahman11b}. This example lends
credence to the NIR color selection method as a means to identify
candidate clusters. In this work, we use the method of
\citet{rahman11a} to identify candidate clusters and OB association
within the remaining SFCs from RM10.

\section{The Selective Star Count Method}
\label{sect:data}

We refine the method of \citet{rahman11a} to identify candidate
clusters and associations within the SFCs using the line-of-sight
extinction through the Galactic plane. This filters non-cluster stars
from candidate members, allowing the associations to be identified in
areas of significant point source contamination. Schematically, the
process is as follows: We isolate stars matching the NIR colors of
O-type stars at graduated extinction levels. We then use the on-sky
density of point sources matching the color cut to find overdensities
consistent with the location of the SFC. Verifying from binned
color-color and color-magnitude diagrams that the candidate cluster is
not caused by a chance overdensity of projected Galactic structure
based on the color distribution along the line of sight, we extract
the candidate cluster's membership properties. We refer to this method
as the Selective Star Count (SSC) method, which we describe in greater
detail.

\subsection{The 2MASS Point Source Catalog}
\label{subsect:2mass}

To identify members of the candidate cluster, we use the 2MASS Point
Source Catalog \citep{skrutskie06}. The survey covers the entire sky
with \jmag\, (1.25 $\micronm$), \hmag\, (1.65 $\micronm$), and \kmag\,
(2.16 $\micronm$) bands, enabling consistent coverage of the SFCs in
both the northern and southern skies. We limit our searches to the
point sources with photometric quality flags of A, B, or C,
corresponding to a minimum signal-to-noise ratio of 5 for all of the
\jmag, \hmag, and \kmag{} bands. The 2MASS catalogue is complete to
\jmag{} $\le 15.8$, \hmag{} $\le 15.1$, and \kmag{} $\le 14.3$
magnitudes in the absence of confusion. However, the areas we
investigate are highly confused causing the completeness limit to be
up to 1 magnitude brighter. This prevents the detection of less
luminous objects that are expected to be cluster members based on the
stellar initial mass function (IMF).

The typical FWHM of the 2MASS point sources is 2\farcs5 and quoted
magnitudes are measured with apertures of 4\arcsec, which will cause
significant confusion for the most compact clusters or the dense areas
of more diffuse OB associations \citep{skrutskie06}. Given the
selected quality limit, these densely-packed confused sources will be
excluded from our searches. Based on this constraint, we expect our
search method to be insensitive to clusters with on-sky point source
densities $\gtrsim 31$ arcminute$^{-2}$ due to overlapping
apertures causing confusion. Within $1\degree$ of the Galactic
midplane, we find a total density of point sources in the target areas
to be between 14 and 17 arcminute$^{-2}$ meeting the chosen
quality cut. Therefore we expect such compact clusters to be readily
identified by visual inspection of the images. In this sense, our SSC
method is entirely complementary to search strategies of
\citet{bica03}, \citet{dutra03} and \citet{mercer05}.  We discuss this
further in the case of NGC 3603 in \S\ref{sect:ngc3603}.

\subsection{Extinction-derived Color Selection}

The hypothesized powering clusters and associations of the SFCs should
exist at large extinctions due to their distance through the Galactic
disk.  We make use of this to identify the cluster candidates on the
basis that all cluster members should be similarly extinguished along
the line of sight. Our method for uncovering candidate clusters is to
identify overdensities of stars with colors consistent with massive
stars at extinction ranges consistent with the distances to the SFC
being probed.

All of the SFCs were identified using techniques sensitive to ionizing
flux with large ionizing luminosities ($Q_{0} \gtrsim
10^{50.5}$\,s$^{-1}$). The most luminous stars near the stellar upper
mass limit have significantly smaller ionizing outputs \citep[$Q_{0} =
10^{49.6}$ s$^{-1}$ for an O3V star; ][]{martins05}. The measured
ionizing luminosities therefore require very large numbers of O-type
stars to power them. For this reason we specifically search for
overdensities of O-type stars with magnitudes and colors consistent
with the distances and expected extinctions to the SFCs.

We use the NIR O-type star magnitudes and colors from
\citet{martins06}. We note that the colors of all O-type stars in
\jh{} and \hk{} are constant, with \jh{} $= -0.11$ and \hk{} $=
-0.10$, as expected, since these bands are in the Rayleigh-Jeans tail
at the surface temperatures of O-type stars. These colors are,
however, not unique to O-type stars; B- and early A-type stars are
indistinguishable from O-type stars based on NIR colors alone to the
accuracy of the 2MASS color measurements \citep{rahman11a}, and later
stars are relatively more abundant; there is a roughly 8:1 ratio
between O and B-type stars \citep{kroupa01}. Only through their
magnitudes can we make a distinction between O-, B- and A-type stars,
given a particular distance and extinction.

We take the intrinsic colors of O-type stars and redden them to arrive
at our 2MASS color cuts. For the extinction law, we adopt that of
\citet{nishiyama09}, which gives the extinction ratios
$\frac{A_{J}}{A_{K}} = 2.89$ and $\frac{A_{H}}{A_{K}}= 1.62$. We note
that this is a departure from the canonical values for the extinction
ratios due to a different power law slope for the relationship between
extinction and wavelength ($A_{\lambda} \propto \lambda^{\alpha} $);
this determination uses a slope $\alpha = -2.0$, while the previously
accepted value is $\alpha = -1.61$ \citep{ccm89}. We adopt the
Nishiyama relationship over the canonical as it is empirically
determined from a more highly extinguished sample of stars that are
more likely to be consistent with the extinctions that we probe in
this work. The shallower slope has been found to be erroneous by
\citet{stead09}, who use an even steeper power law slope of $\alpha =
-2.14$.

The steeper slope of the extinction law changes the relationship
between the NIR ($> 1\,\mu$m) and visual ($\sim 0.5\,\mu$m); rather
than a ratio of $A_{V}/A_{K} \sim 10$, the new ratio becomes
$A_{V}/A_{K} \sim 16$ \citep{nishiyama08}. However, the difference in
the extinction ratios between the different laws in the NIR ($1 -
2.5\,\mu$m) bands is less than 20\%, which is inconsequential in the
candidate cluster identification since we use only the NIR bands and
the large color cuts used for the selection procedure.

We use the following relations for $A_{K}^\prime$, the estimated
O-star extinction in the K-band:
\begin{eqnarray}
A_K^\prime(J-H) & = & 0.797(J-H) + 0.0877 \\
A_K^\prime(H-K_S) & = & 1.66(H-K_S) + 0.166
\end{eqnarray}

We use these relations to select point sources consistent with O-type
stars in a particular extinction range.

\subsection{The Nearest Neighbour Algorithm}

We analyze the density structure of 2MASS point sources using a
nearest neighbour method, as described by \citet{casertano85}. This
method samples each of the points within the chosen color cut,
providing an adaptive resolution, sampling the areas with more point
sources more heavily than those areas without point sources. The
primary advantage of this method over a simple spatial density map is
that the resolution is adaptive; areas with large numbers of point
sources will contain detail of density structures at a higher
resolution than areas with a sparsity of sources. From each of the
points, we measure the distance $r_j$ to the $j$-th nearest
neighbour. We determine the local surface density, $\mu_{j}$, with an
unbiased estimator:
\begin{equation}
  \mu_{j} = \frac{j-1}{\pi r_j^2}
\end{equation}      
with $j > 3$. From \citet{casertano85}, the standard deviation of the
estimator is:
\begin{equation}
  \sigma_{j} = \frac{\mu_{j}}{\sqrt{j-2}}
\end{equation}

To ensure the consistent visual appearance of the point source density
variations regardless of the number of sources in the field, the
specific value of $j$ to be used is a function of the total number of
point sources in the field. If a constant $j$ value was used for all
fields, those with large numbers of point sources would sample only
very local variations of the surface density, while those with few
point sources would smooth over all but the largest density structures
in the field. The on-sky point source density varies based on the
location in the Galactic plane, the color-cut chosen, and the width of
the color-cut. The typical on-sky densities of the point sources can
vary by an order of magnitude based on different color cuts since
fewer stars with very red colors are cataloged in 2MASS than bluer
stars. Consequently, using a constant $j$ value over all color cuts
would dramatically change the scale of structure highlighted. So for
each cut, we determine an optimal value for $j$ given a angular
resolution based on the average number of point sources within a
field:

\begin{equation}
j = \lfloor \frac{\theta_{cr}^{2} N_{*}}{\Theta^2} \rfloor
\end{equation}
where $N_{*}$ and $\Theta$ are the total number of stars in and
angular radius of the color-selected 2MASS field, and $\theta_{cr}$
is the {\it characteristic resolution}, corresponding to the mean
resolution across the selected field. For the purpose of this
investigation, we choose a characteristic resolution $\theta_{cr} =
3.5'$ based on the resolution used to identify the Dragonfish
Association from \citet{rahman11a}. At this characteristic resolution,
typical $j$ values range from 15 to 30 for most regions with
extinction cuts of $\Delta$A$_K = 0.1$. The densities are gridded onto
an oversampled ($d < 11''$), uniformly-spaced grid using a Delaunay
triangulation method from the Matplotlib python package
\citep{hunter07} to simplify the analysis.

\subsection{Candidate Identification}
\label{sect:candid}
For each of the SFCs, we examine the surrounding 2\degr{} field with
0.1 steps in the A$_{K}^\prime$ color cut in the range of
$0.0<$A$_{K}^\prime<2.0$. We enlarge the surrounding field to 3\degr{}
for SFCs with semi-major axes larger than 30\arcmin. In each
extinction step, we produce the nearest neighbour on-sky density map,
and visually identify cluster candidates coincident with the on-sky
position of the SFC position and the 8 \micron{} morphology from RM10.

Using small steps in the color cut allows for effective filtering of
field stars from candidate clusters: while the typical point source
density of unfiltered sources ranges from 14 to 17 sources
arcminute$^{-2}$ (\S\ref{subsect:2mass}), the typical field densities
within the color cuts range from 0.05--0.5 sources
arcminute$^{-2}$. The substantial decrease in the field source density
due to the color filtering enables the detection of more diffuse
clustering than has been possible in previous searches
(\S\ref{sect:prev}).

\begin{figure*}
\begin{center}
\includegraphics[scale=0.75]{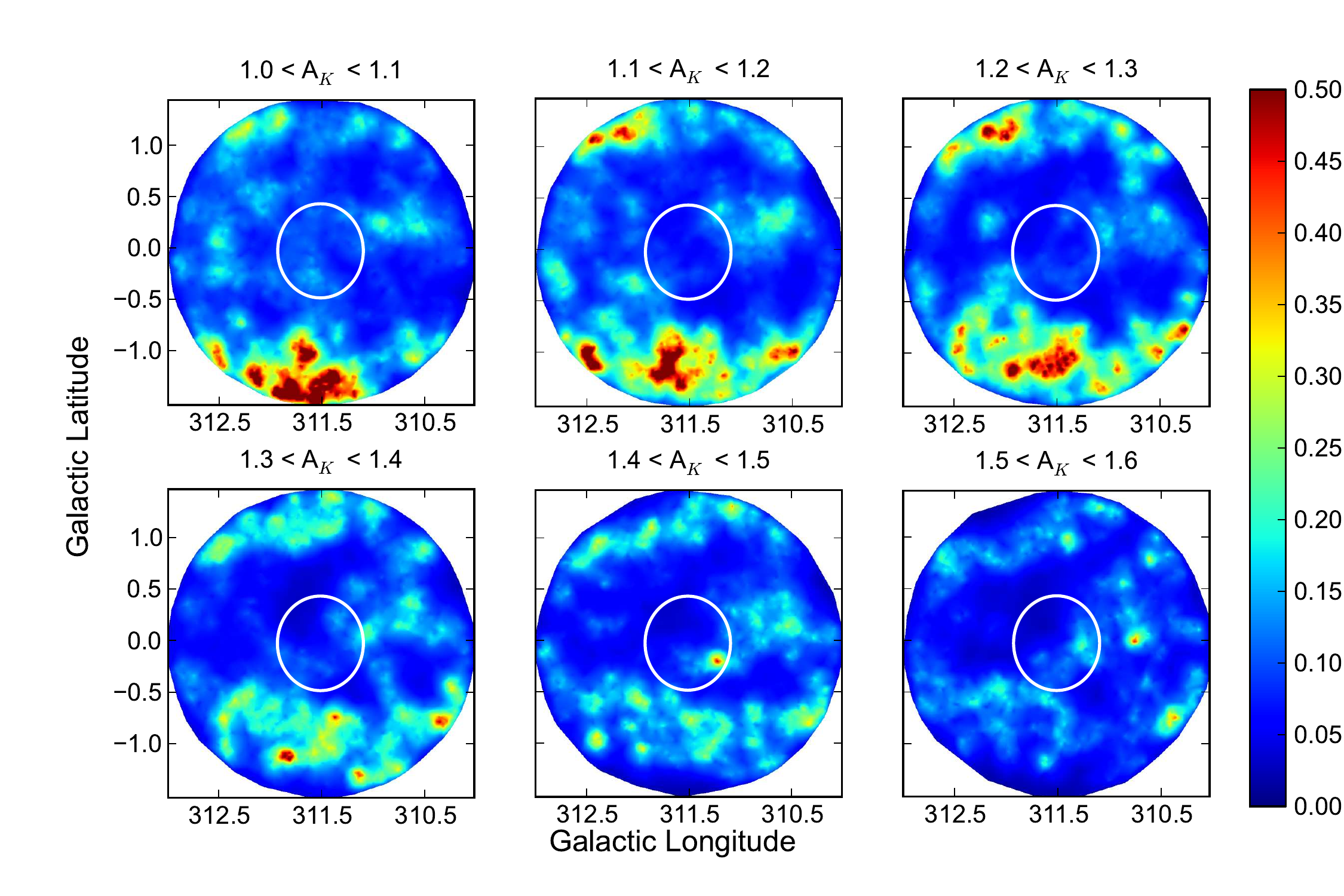}
\end{center}
\caption{2MASS Point Source Density Maps at the location of SFC 31
  (white circle) with extinction cuts between A$_{K}$ = 1.0 and 1.6
  (left to right, top to bottom). The color bar indicates the point
  source density in sources arcminute$^{-2}$. Galactic structure
  effects produce the long horizontal streak-like features that appear
  in the first extinction cut symmetrically 1\degr{} above and below
  the galactic plane and narrow in subsequent cuts. These features are
  visually filtered from the selection process. Despite the appearance
  of compact overdensities, these cuts do not contain any identified
  candidate clusters, since they are likely higher-density pockets
  associated with the Galactic structure features. This can be
  compared to the candidate identified in this region at a lower
  extinction cut (Table \ref{tab:cclist}; Figure
  \ref{fig:sfc283031}). \label{fig:galstruct}}
\end{figure*}

Our adoption of sequential extinction cuts over the entire range
allows for the filtering of effects due to the structure of the Galaxy
itself. For instance, a significant effect of Galactic structure is
the appearance of patchy but mostly latitude-dependent stratification
of the source density above and below the Galactic plane caused by the
limited thickness of the extinction-producing gas disk and also by the
appearance of all thick disk and halo stars at a particular color
range (Figure \ref{fig:galstruct}). These large scale effects can be
visually separated from localized overdensities with ease.

Once a candidate is identified, the extinction range is refined to
enhance the prominence of the overdensity, and a final density map is
produced. From this map we visually identify the candidate's geometric
boundaries, including its centroid position, semi-major and semi-minor
axes. Once the boundaries are chosen, we determine the total number of
enclosed sources, $N_{TOT}$. To constrain field contamination, we
measure the mean ($N_{BG}$) and standard deviation ($\sigma_{BG}$) of
the number of field stars surrounding the candidate in regions with
the same area as the identified candidate.

To better distinguish between real clusters and asterisms, we analyze
each candidate cluster with binned color-color and color-magnitude
diagrams as in \citet{rahman11a}. Both sets of diagrams allow us to
distinguish between a true overdensity of sources (evidence for the
existence of a cluster or association) or a shift of sources from one
color to another (evidence of a local extinction feature). From these
diagnostics, we find that none of the overdensities appear to be
shifts of points in color-color or color-magnitude space, which is
additional evidence against the alternative hypothesis of a local
extinction feature.

\section{Candidate Clusters/Associations}
\label{sect:candclust}

In Table \ref{tab:cclist}, we present the candidate
clusters/associations identified within the SFCs. For SFCs with
multiple candidates, an alphabetical key is used to distinguish
between them. To quantitatively measure the statistical significance
of the extracted clusters as compared to the natural field variations,
we define the extraction significance:

\begin{equation}
{\rm ES} = \frac{N_{TOT} - N_{BG}}{\sigma_{BG}}
\end{equation}
with $N_{TOT}$, $N_{BG}$, and $\sigma_{BG}$ as defined in
\S\ref{sect:candid}.

Two of the candidate associations (those within SFC 8 and 30) have
extraction significance values of less than unity; these candidates
are identified solely based on their morphology as compared to the SFC
and have significant large scale structure in the field surrounding
them that produces large background variations.

The candidates have characteristic radii between 3\arcmin{} and
20\arcmin, with a median of $\sim 7\arcmin$. This is consistent with
the sensitivity of the SSC method to larger diffuse structures than in
previous cluster searches (see \S \ref{sect:candid}). Accordingly,
these candidates would not have been identified in the previous
searches. While the SSC method is more sensitive to diffuse
structures, the ellipses are chosen to contain the most prominent
areas of the overdensity and do not necessarily represent the outer
boundaries of the clusters. We expect that K-giants are the primary
contaminants of all the new candidates, since they tend to be the
dominant population of visible red stars in the 2MASS survey
\citep{robin03}.

The candidate boundaries are visually chosen to encompass the most
centrally-condensed, smooth portions of the overdensity, avoiding the
lower isodensity contours, which are more flocculent, when
possible. This corresponds to an extraction contour of 1--$\sigma$ to
2.8--$\sigma$ above the local background variation level. We
investigate any bias in the visual identification of candidates, the
chosen extraction contour, and their effects on their extracted
parameters with a suite of tests described in Section
\ref{subsect:exttest}.

\begin{figure*}
  \begin{center}
    \includegraphics[scale=0.5]{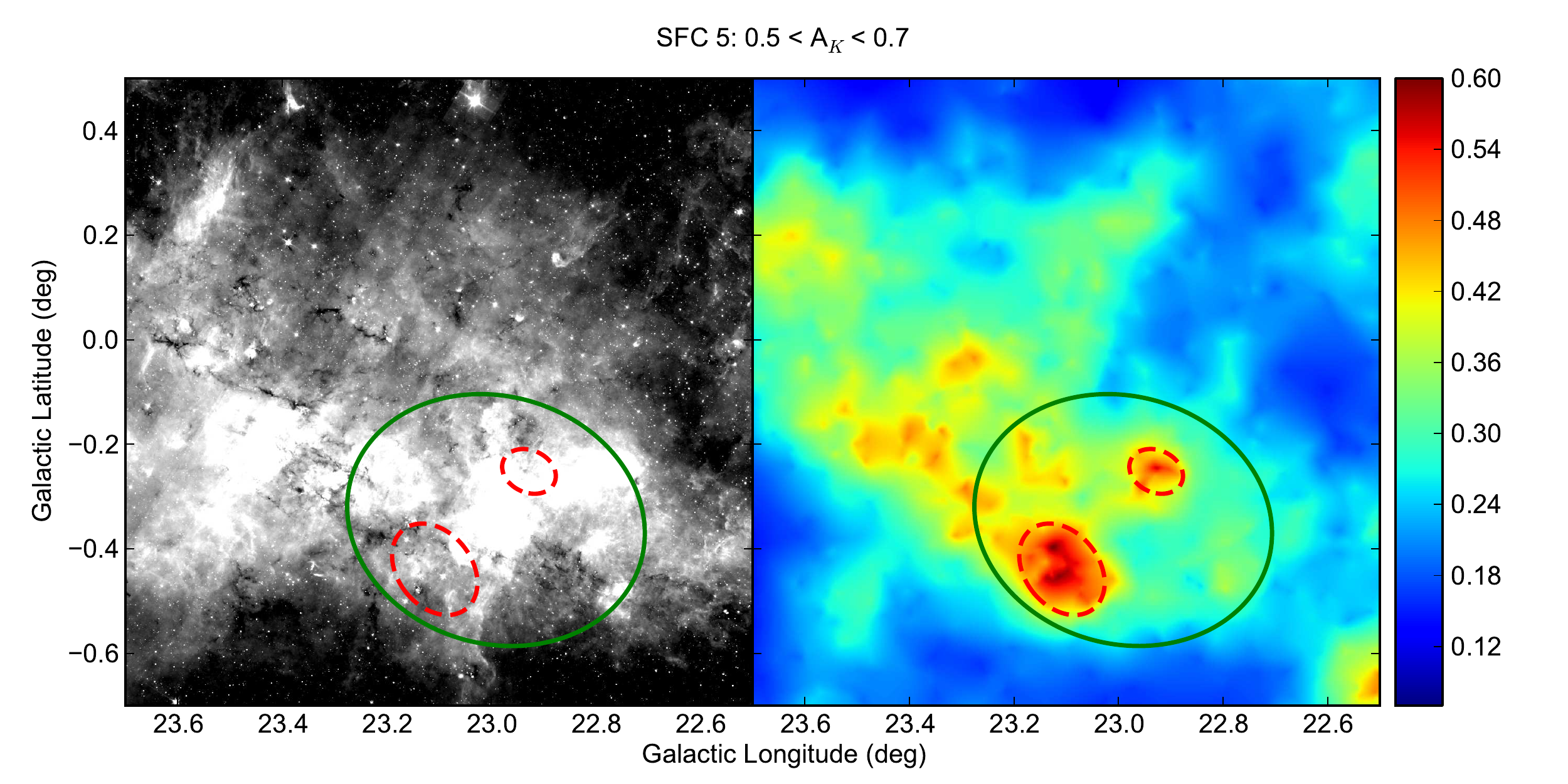}\\
    \includegraphics[scale=0.5]{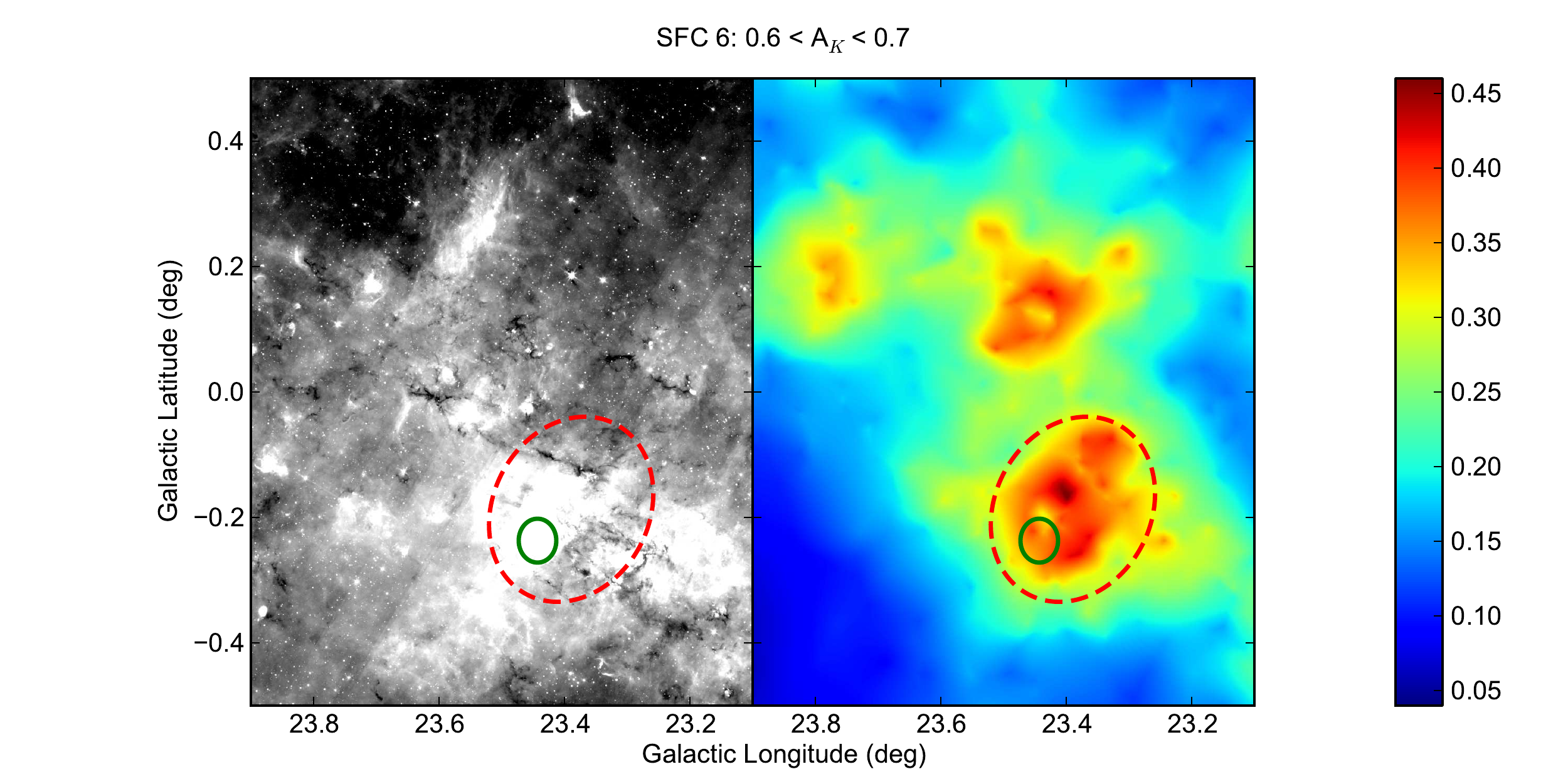}\\
    \includegraphics[scale=0.5]{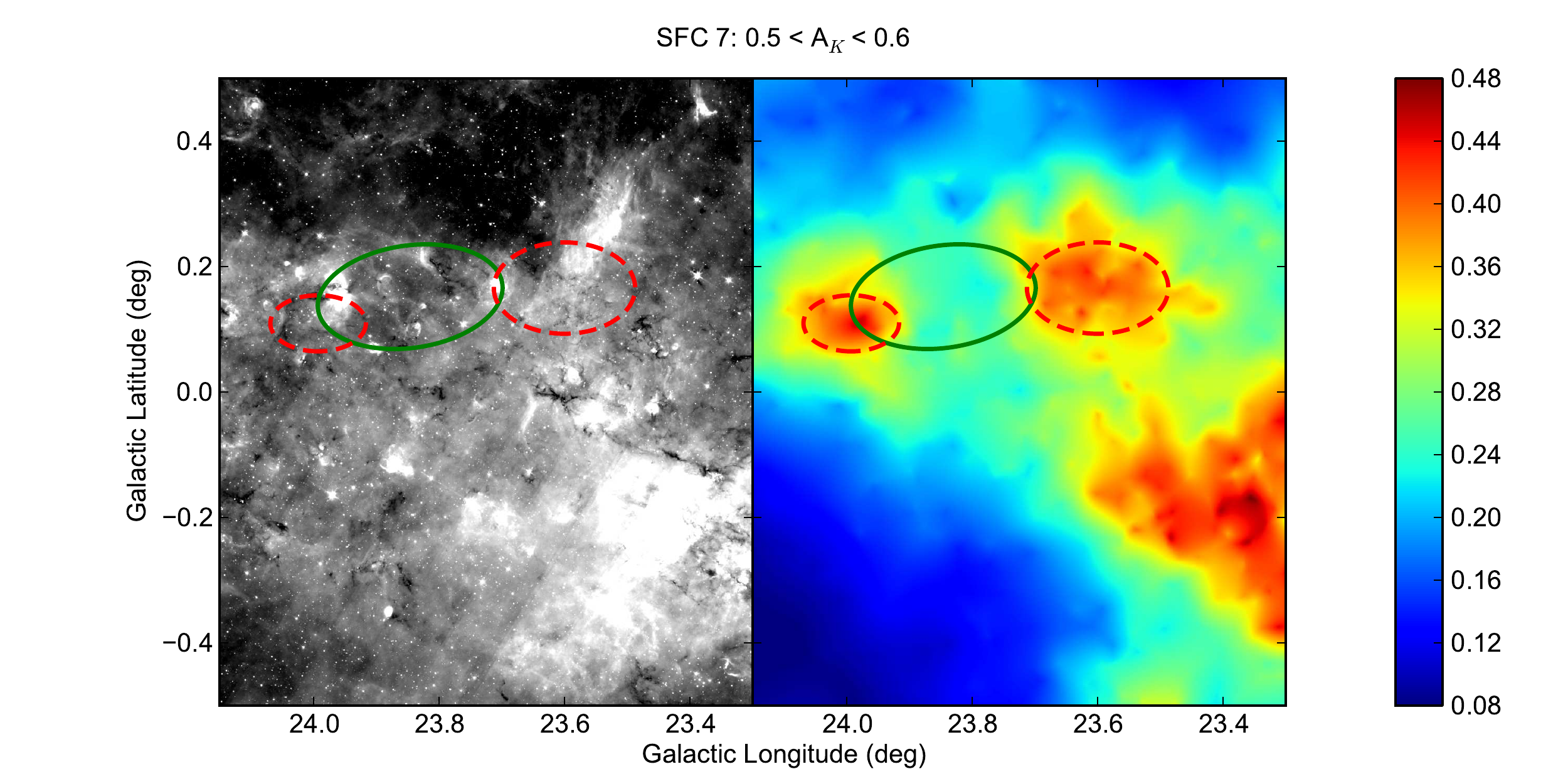}\\

  \end{center}
  \caption{Spitzer GLIMPSE 8$\micronm$ image (left) and the on-sky
    density diagrams (right) of the 2MASS point sources. The source
    and extinction ranges are indicated above each figure. The green
    solid ellipse indicates the location of the SFC from RM10, while
    the red dashed ellipses indicate the location of the candidate
    clusters. The color bar indicates the point source density in
    sources per square arcminute. \label{fig:sfc050607}}
\end{figure*}

\begin{figure*}
  \begin{center}
    \includegraphics[scale=0.5]{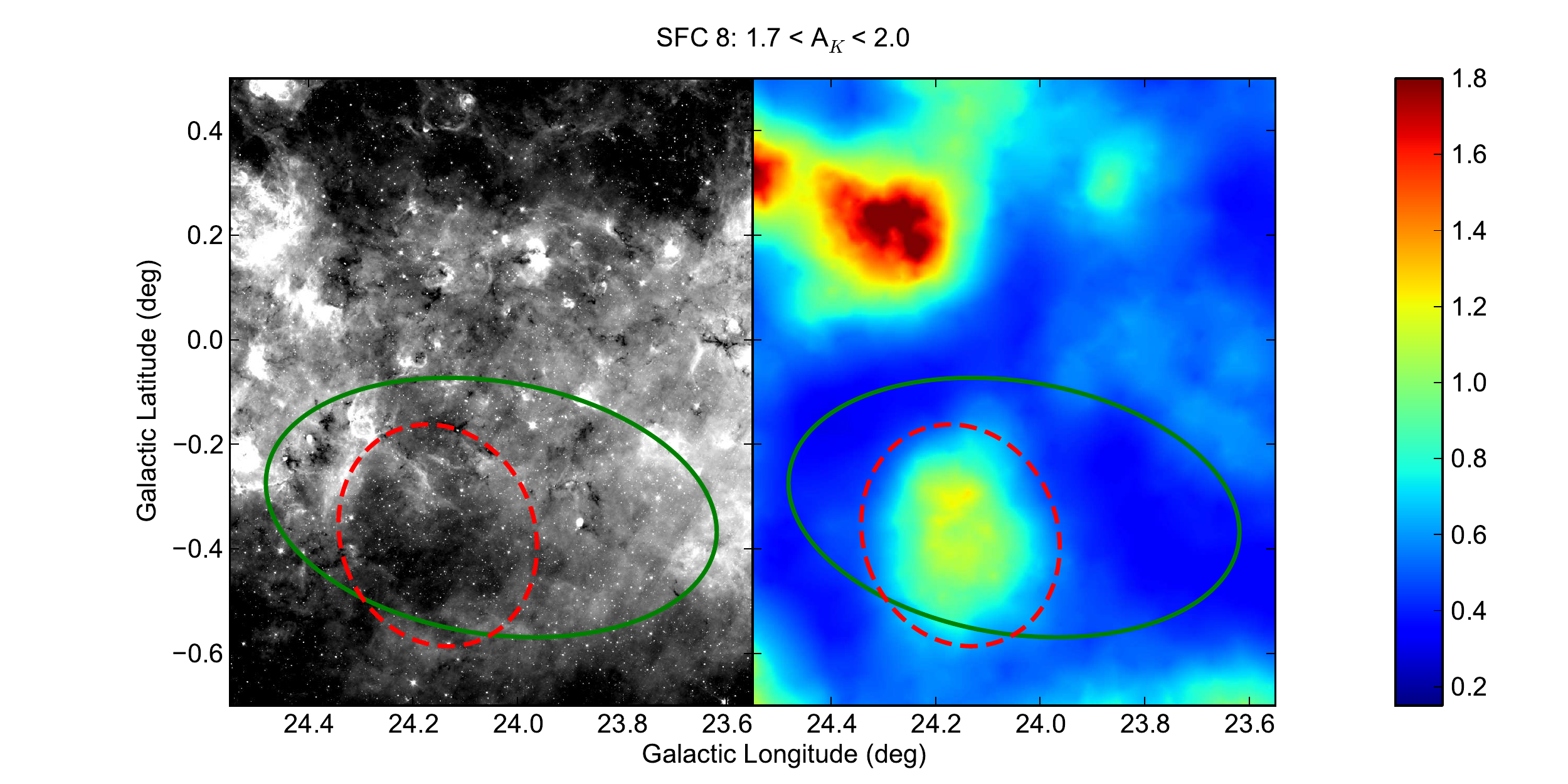}\\
    \includegraphics[scale=0.5]{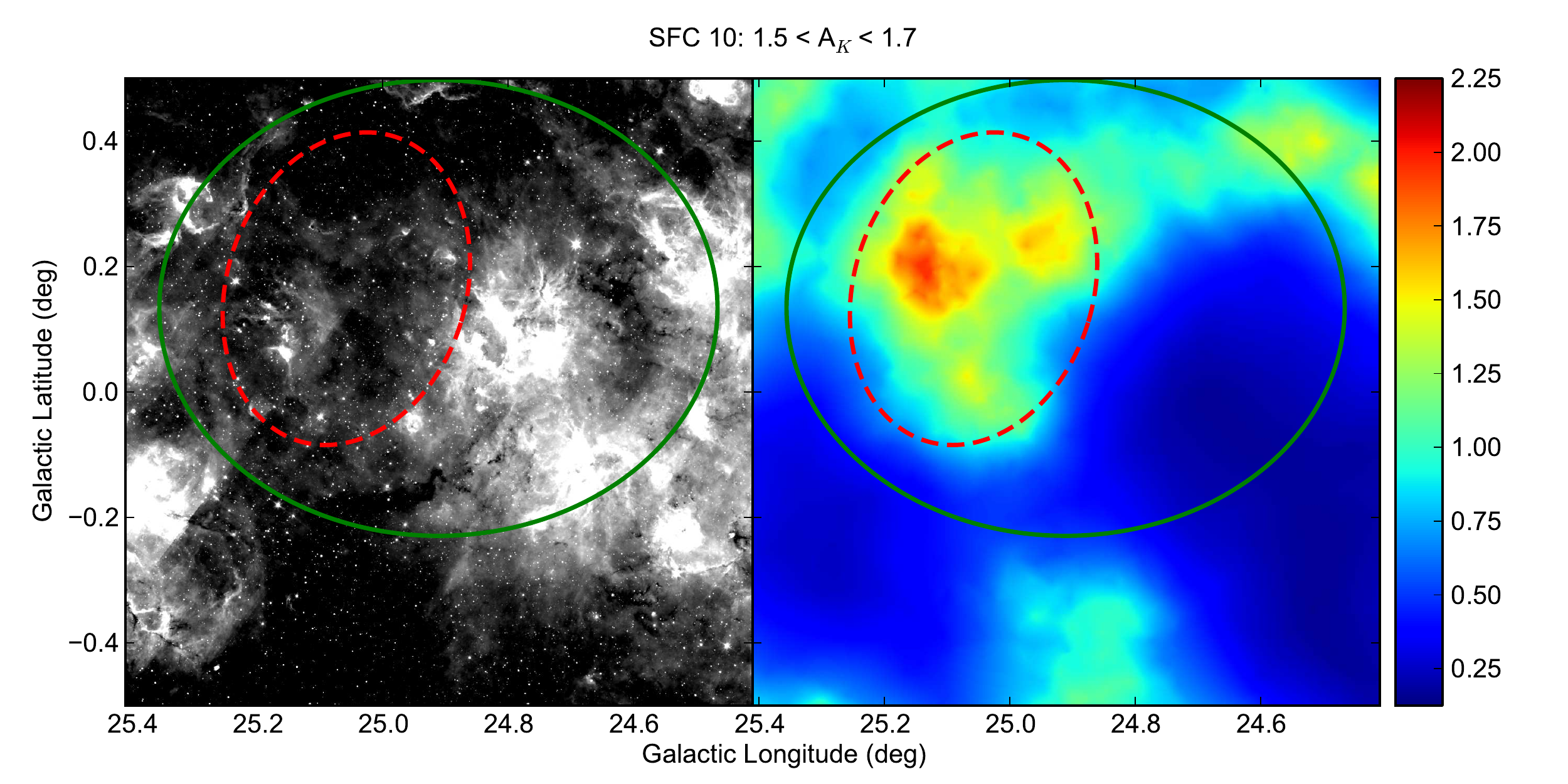}\\
    \includegraphics[scale=0.5]{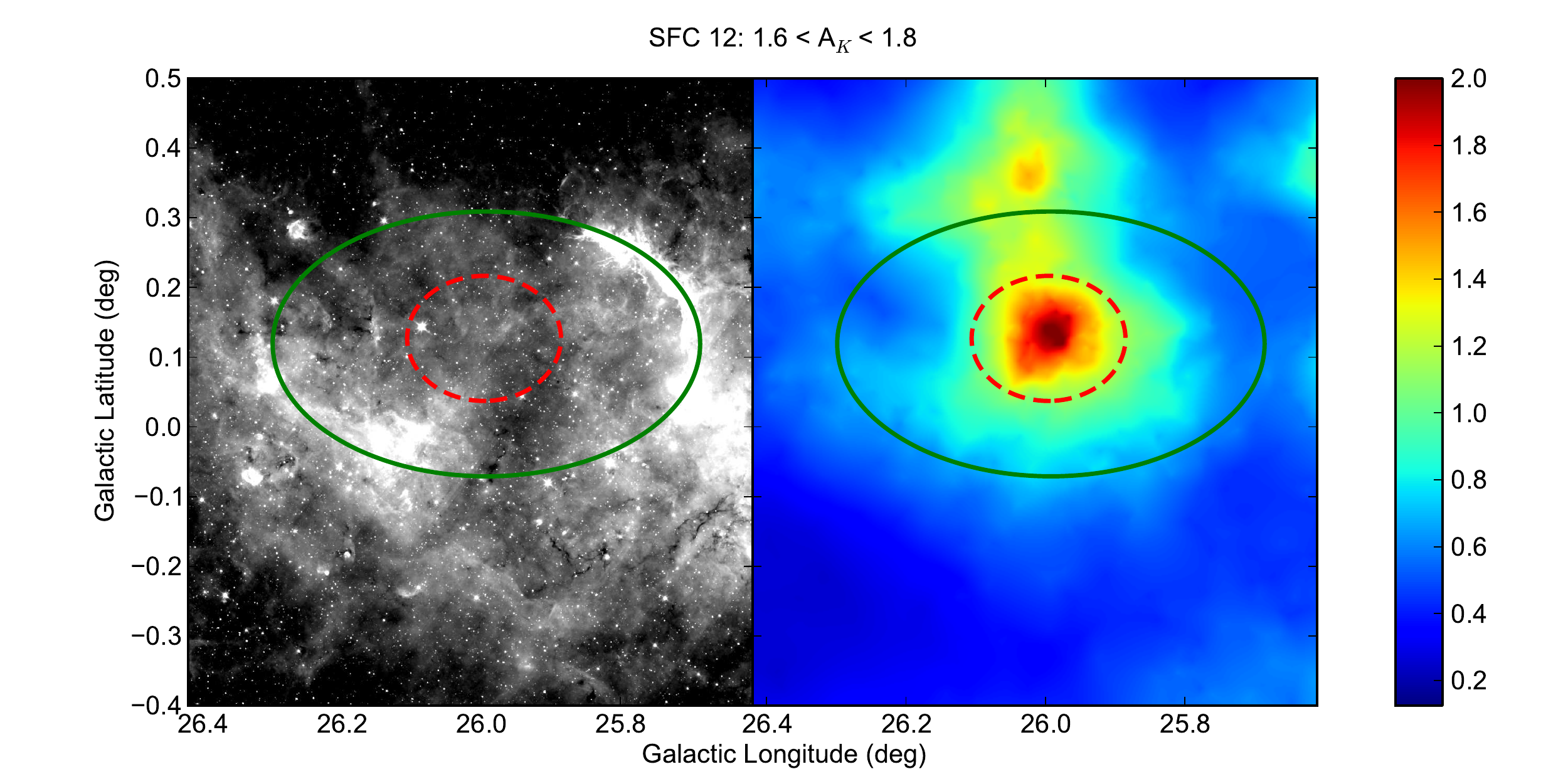}\\
  \end{center}
  \caption{Spitzer GLIMPSE 8$\micronm$ image (left) and the on-sky
    density diagrams (right) of the 2MASS point sources. The source
    and extinction ranges are indicated above each figure. The color
    bar and annotations are the same as Figure
    \ref{fig:sfc050607}. \label{fig:sfc081012}}
\end{figure*}

\begin{figure*}
  \begin{center}
    \includegraphics[scale=0.5]{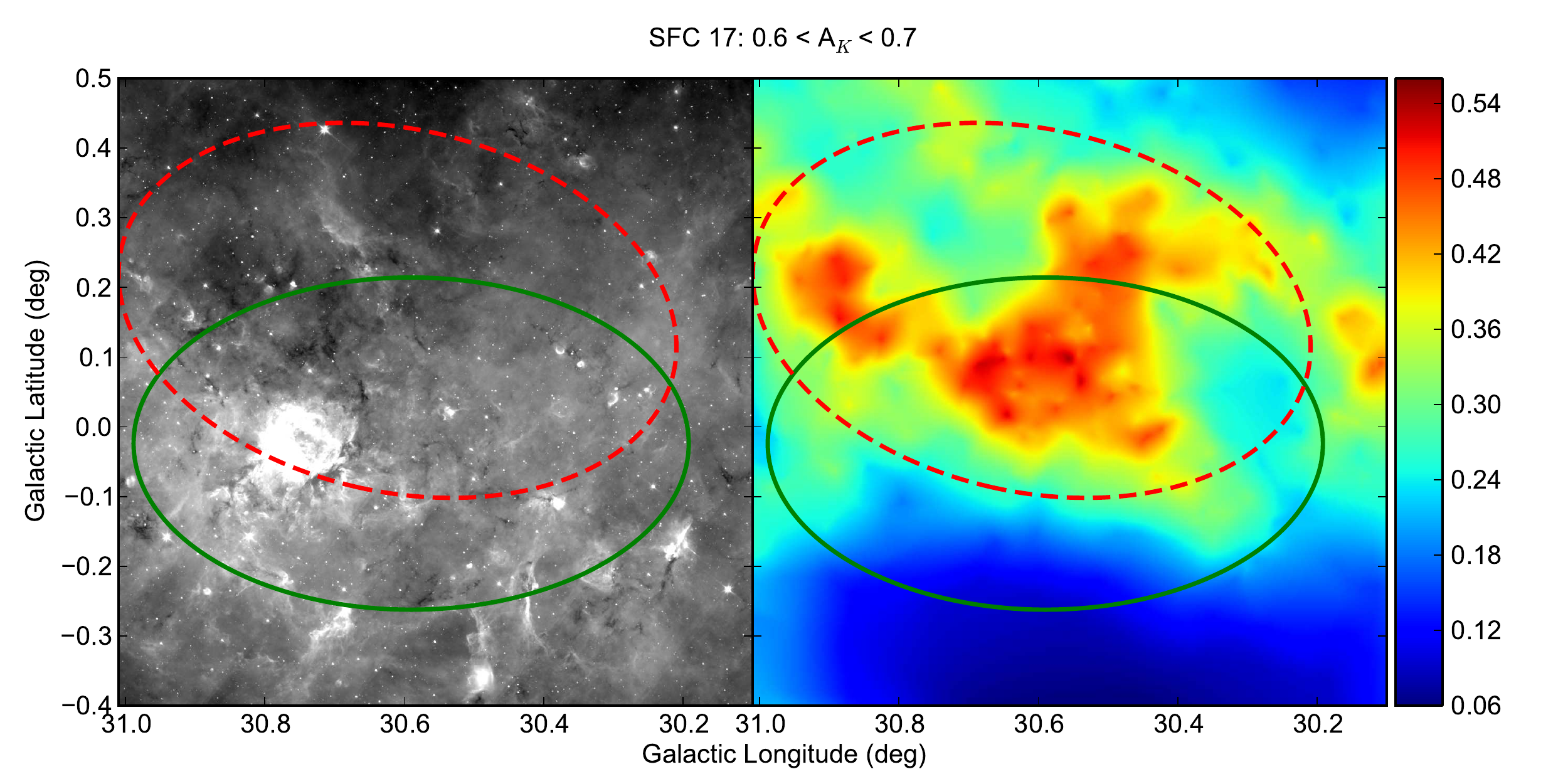}\\
    \includegraphics[scale=0.5]{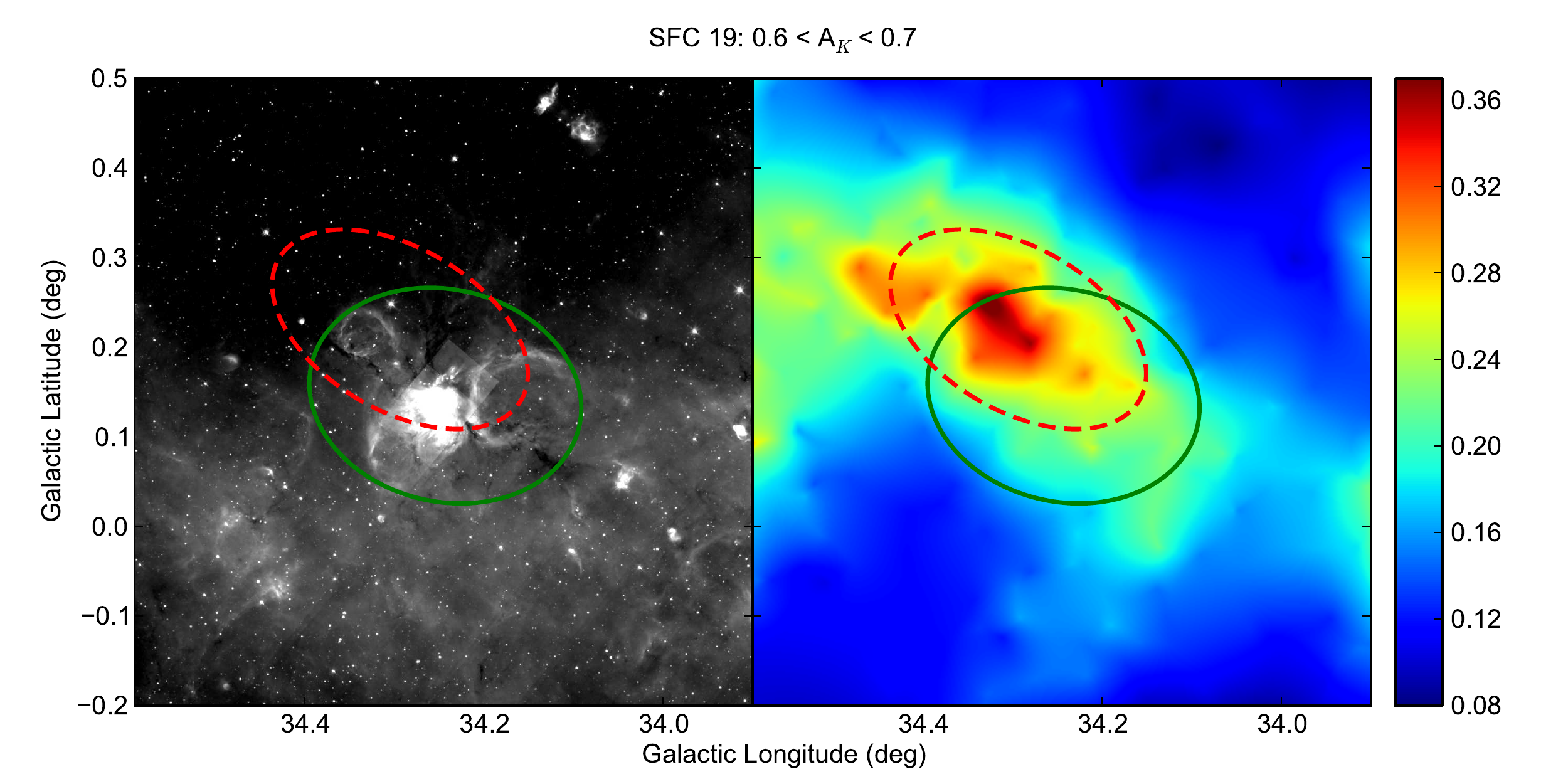}\\
    \includegraphics[scale=0.5]{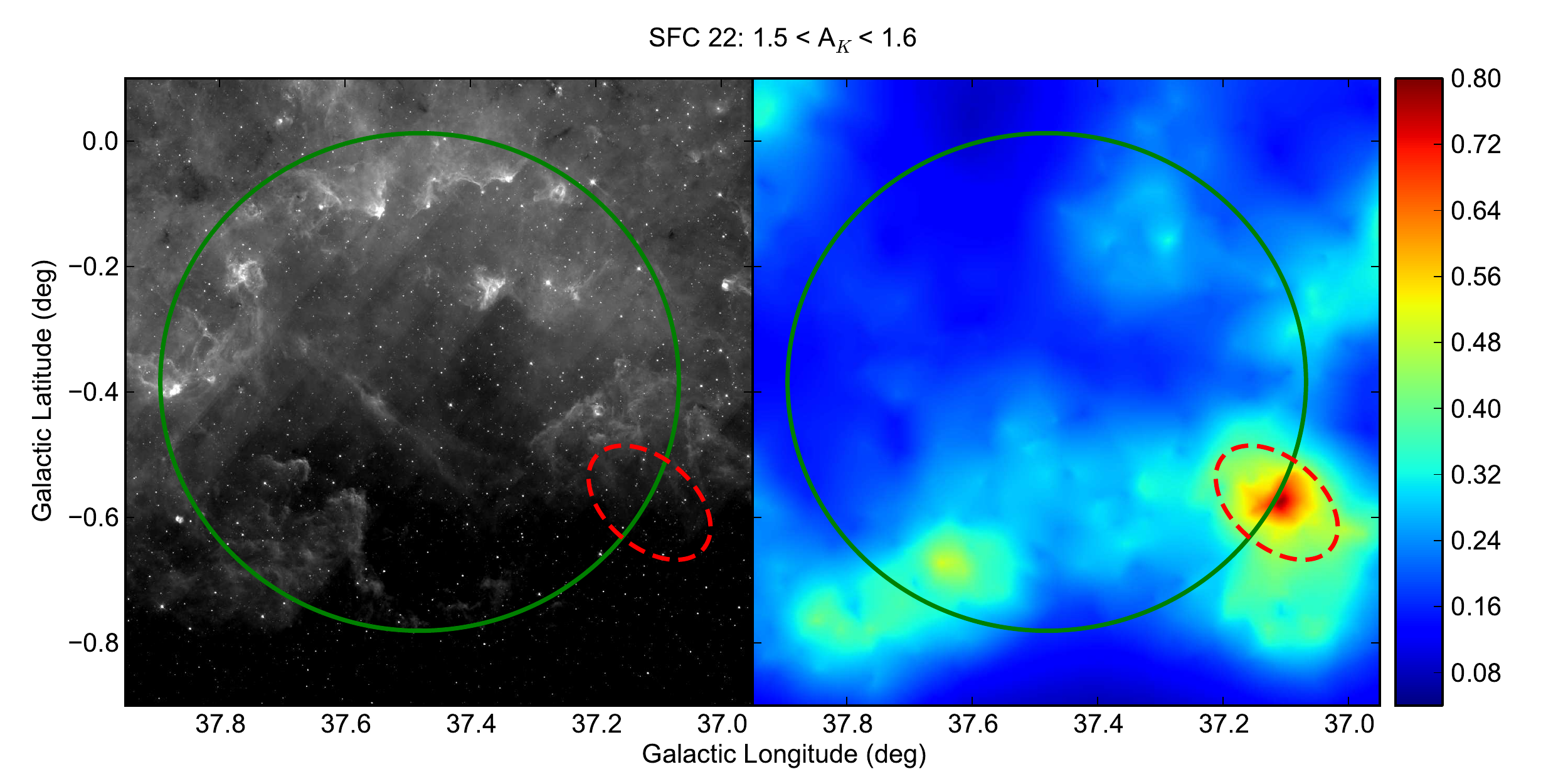}\\
  \end{center}
  \caption{Spitzer GLIMPSE 8$\micronm$ image (left) and the on-sky
    density diagrams (right) of the 2MASS point sources. The source
    and extinction ranges are indicated above each figure. The color
    bar and annotations are the same as Figure
    \ref{fig:sfc050607}. \label{fig:sfc171922}}
\end{figure*}

\begin{figure*}
  \begin{center}
    \includegraphics[scale=0.5]{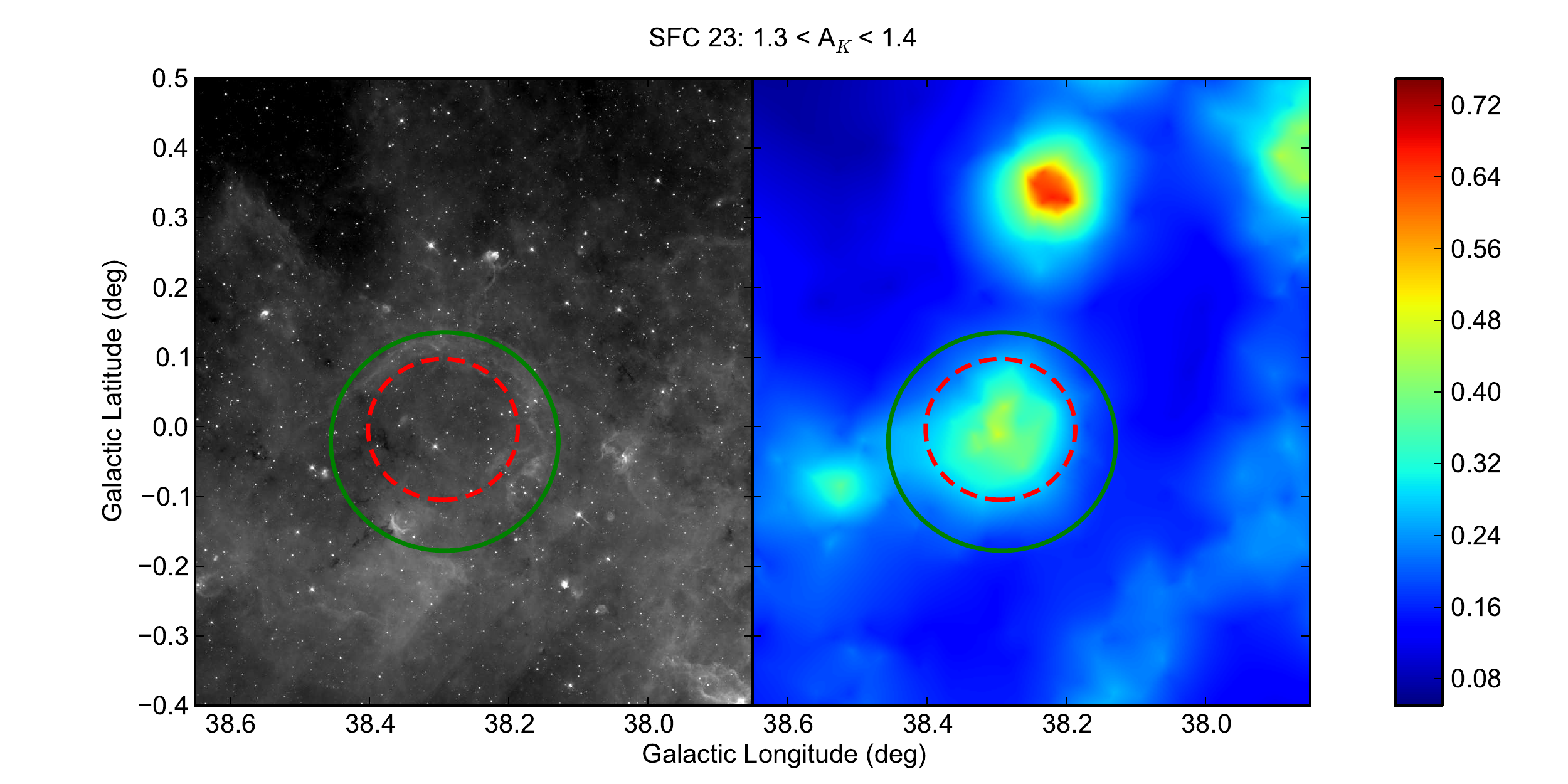}\\
    \includegraphics[scale=0.5]{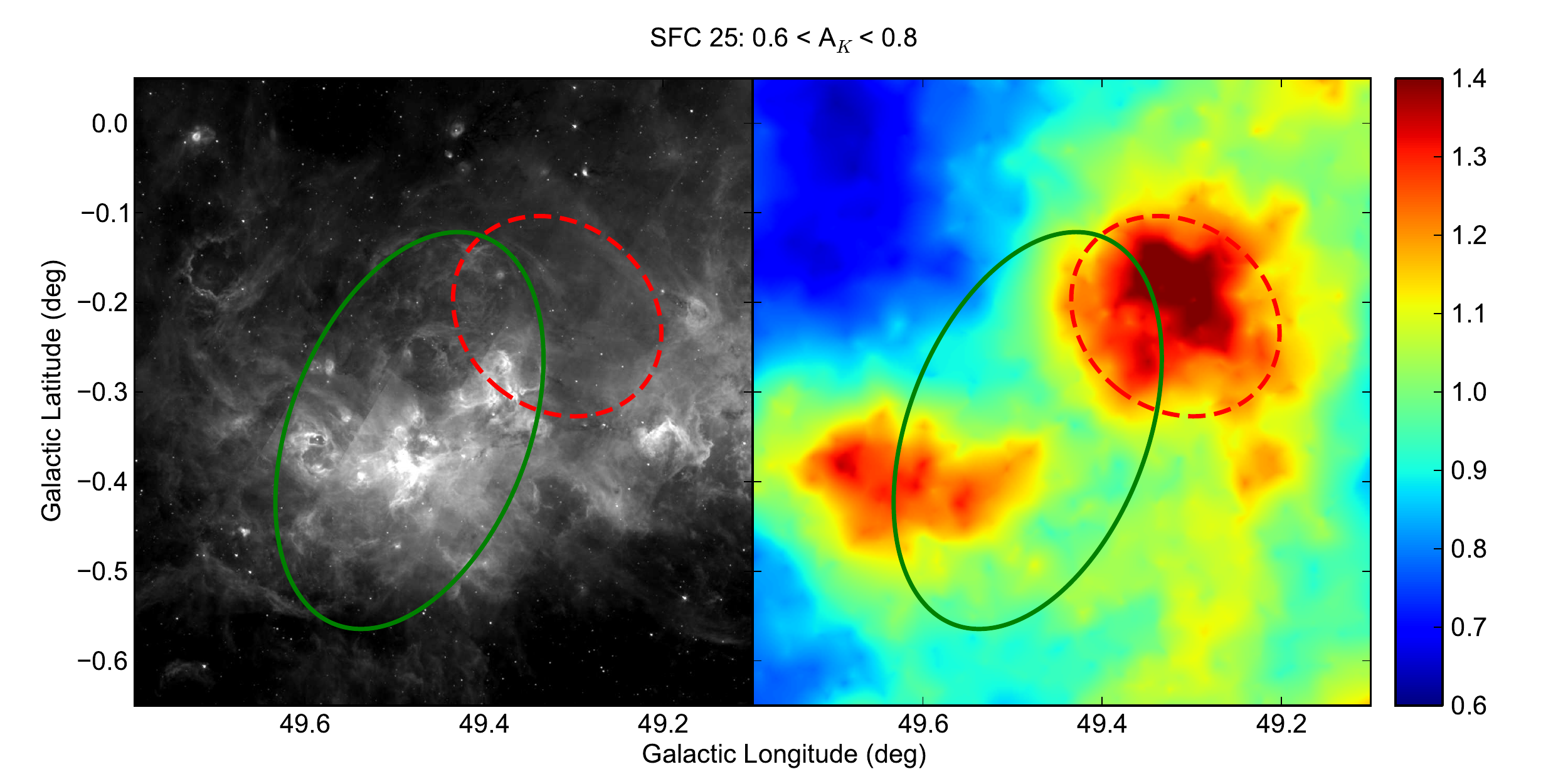}\\
    \includegraphics[scale=0.5]{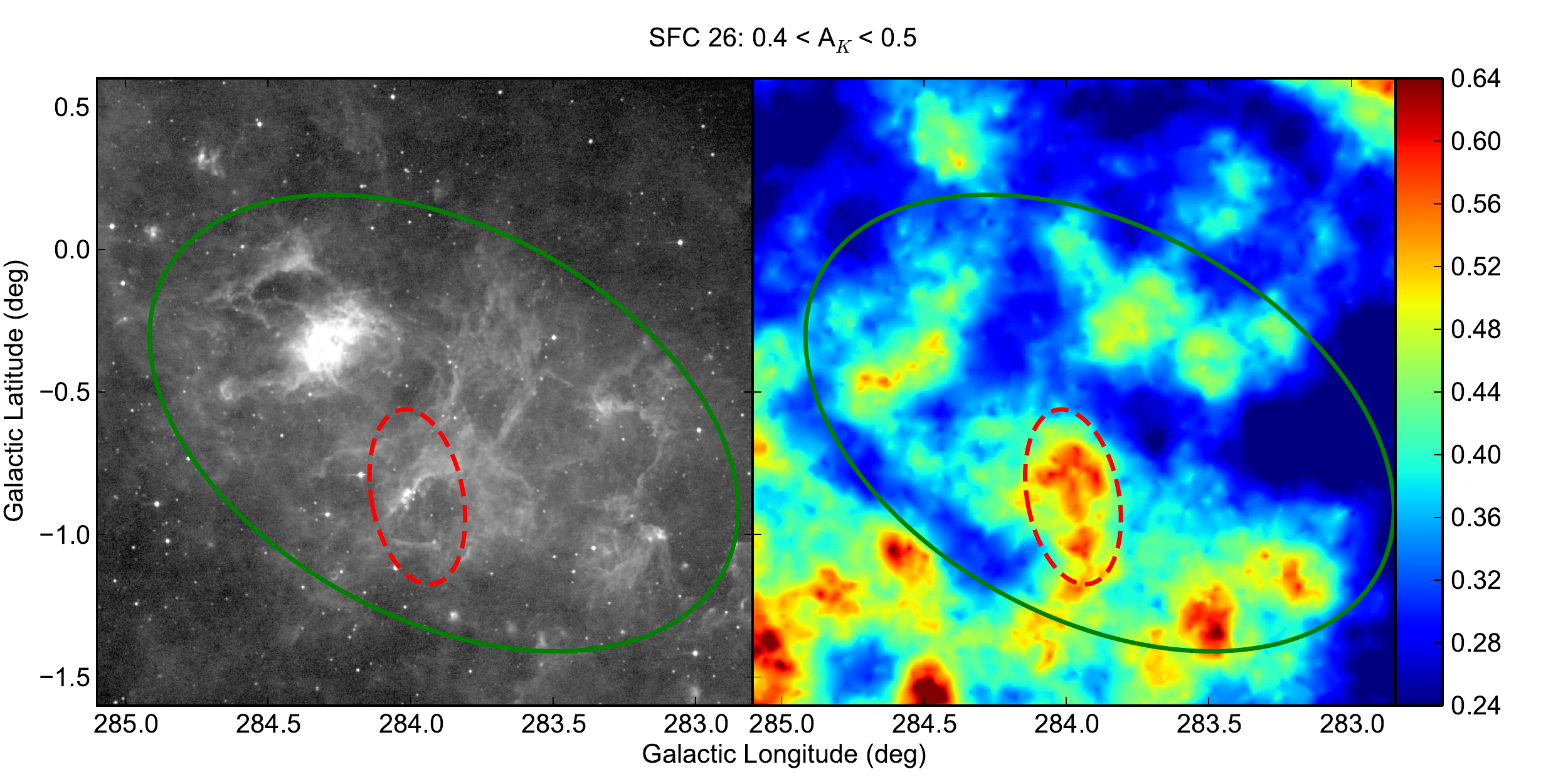}\\
  \end{center}
  \caption{Spitzer GLIMPSE 8$\micronm$ image (left) and the on-sky
    density diagrams (right) of the 2MASS point sources. The source
    and extinction ranges are indicated above each figure. The color
    bar and annotations are the same as Figure \ref{fig:sfc050607}. We
    use the MSX 8 $\micronm$ image for SFC 26 due to Spitzer GLIMPSE
    coverage limits. \label{fig:sfc232526}}
\end{figure*}

\begin{figure*}
  \begin{center}
    \includegraphics[scale=0.5]{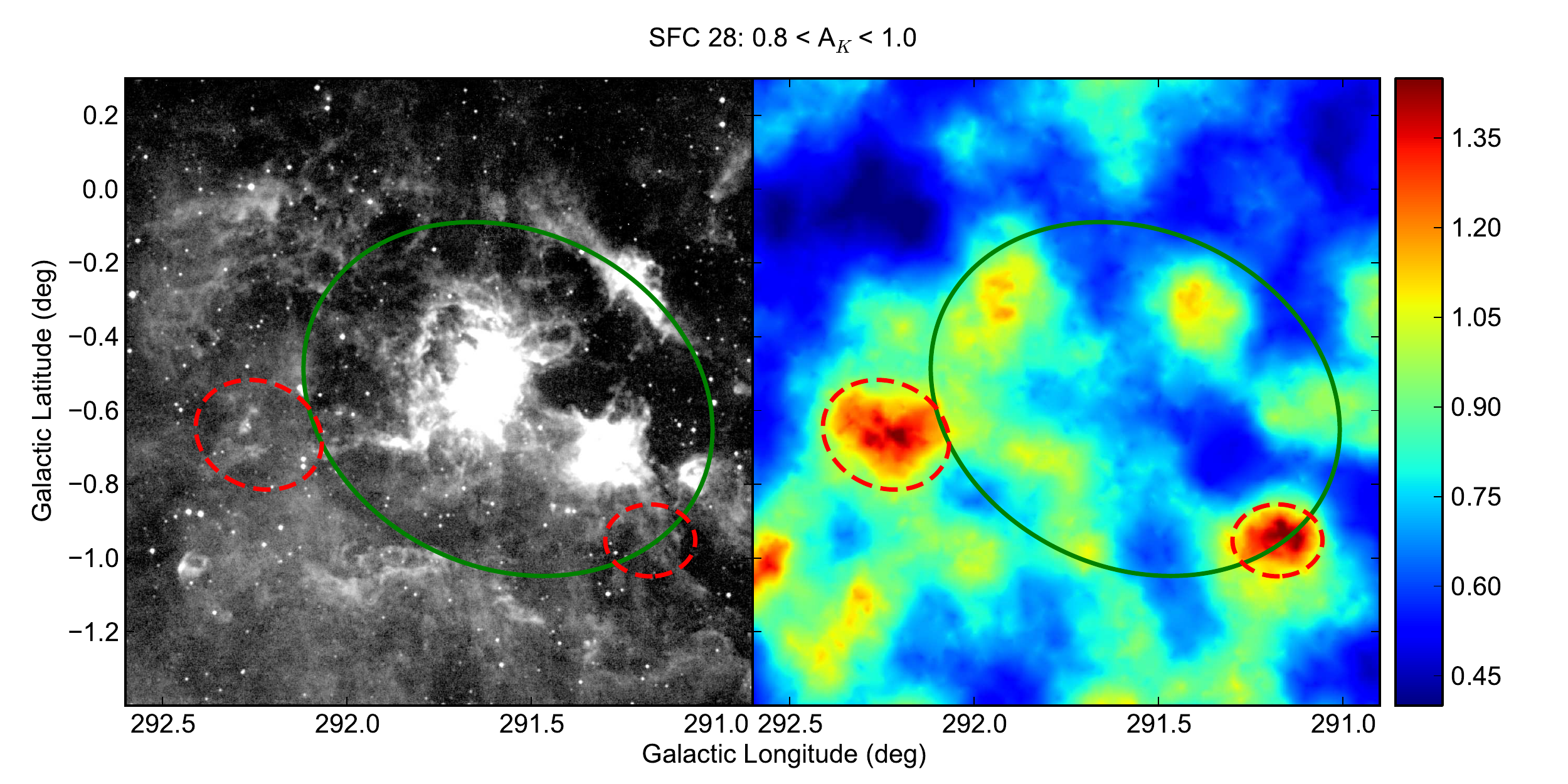}\\
    \includegraphics[scale=0.5]{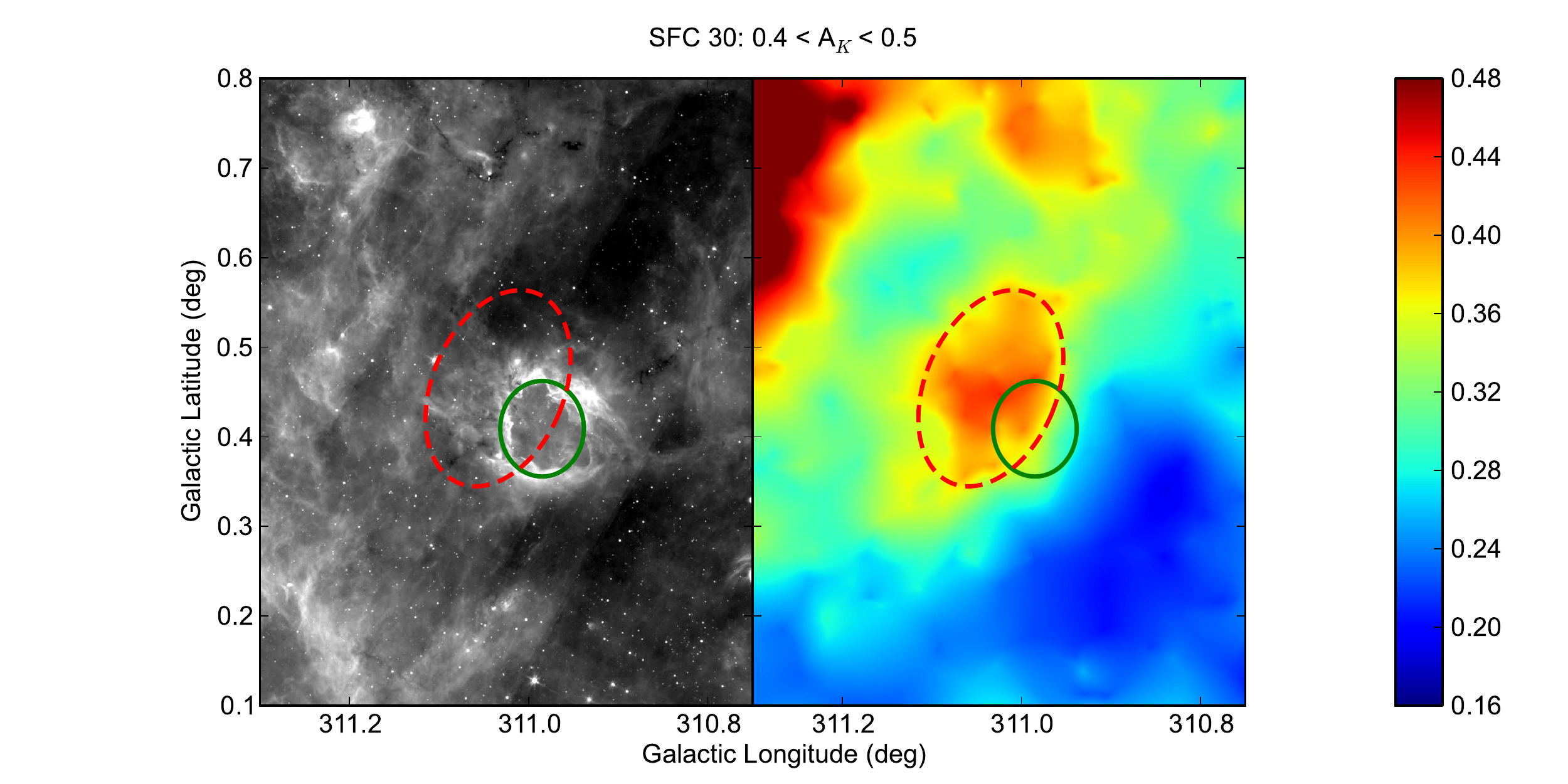}\\
    \includegraphics[scale=0.5]{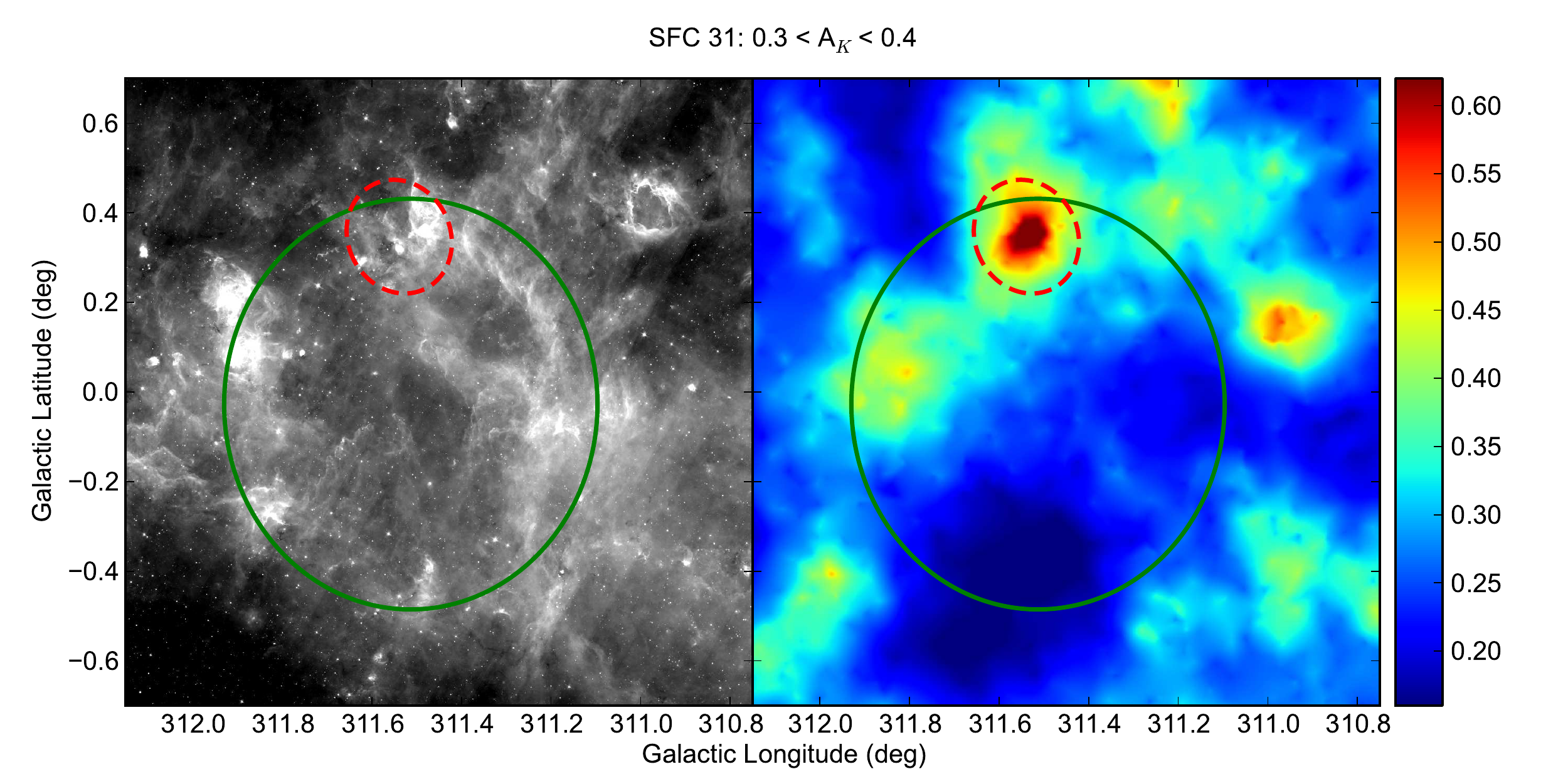}\\
  \end{center}
  \caption{Spitzer GLIMPSE 8$\micronm$ image (left) and the on-sky
    density diagrams (right) of the 2MASS point sources. The source
    and extinction ranges are indicated above each figure. The color
    bar and annotations are the same as Figure \ref{fig:sfc050607}. We
    use the MSX 8 $\micronm$ image for SFC 28 due to Spitzer GLIMPSE
    coverage limits.  \label{fig:sfc283031}}
\end{figure*}

\begin{figure*}
  \begin{center}
    \includegraphics[scale=0.5]{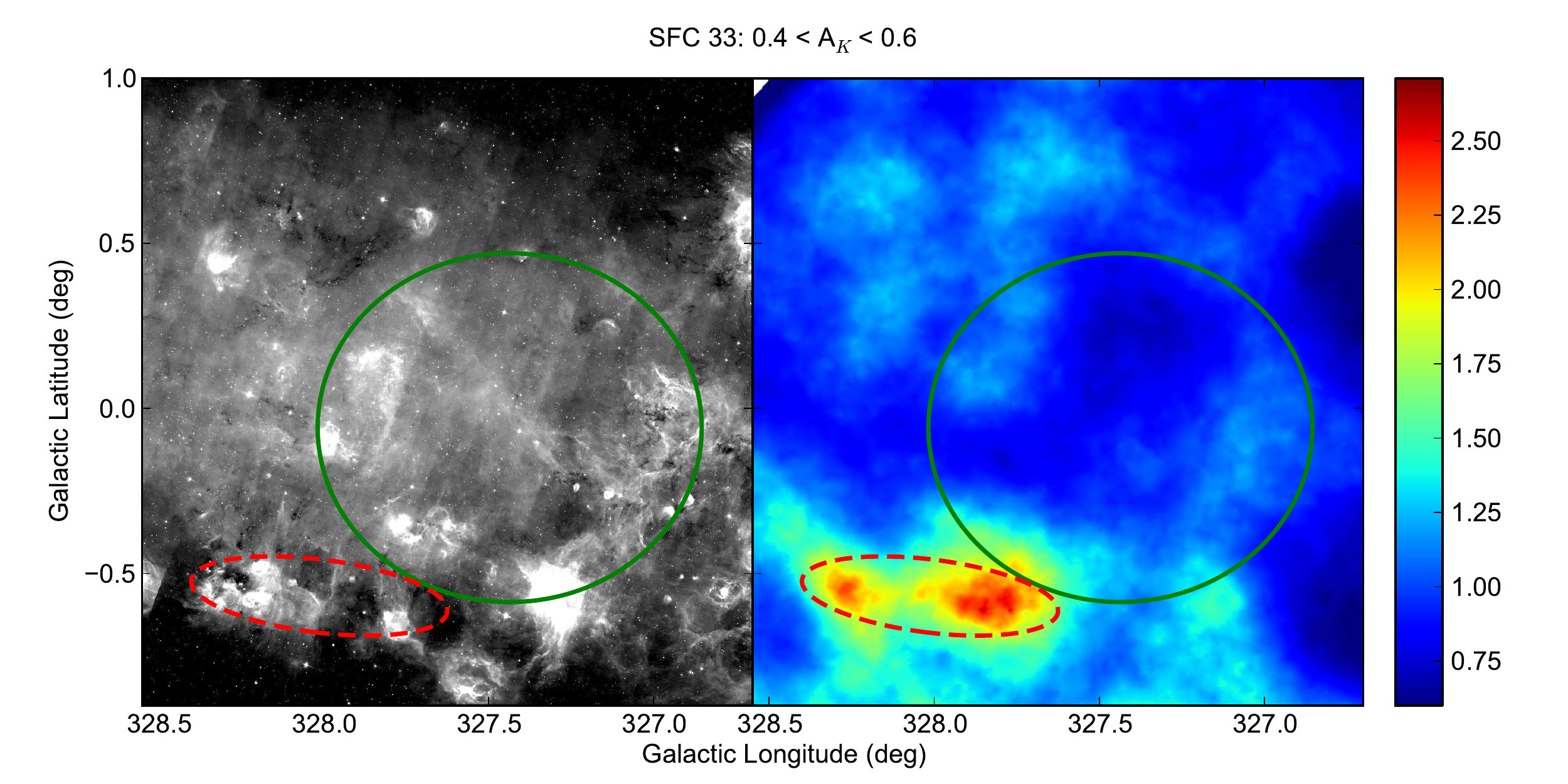}\\
    \includegraphics[scale=0.5]{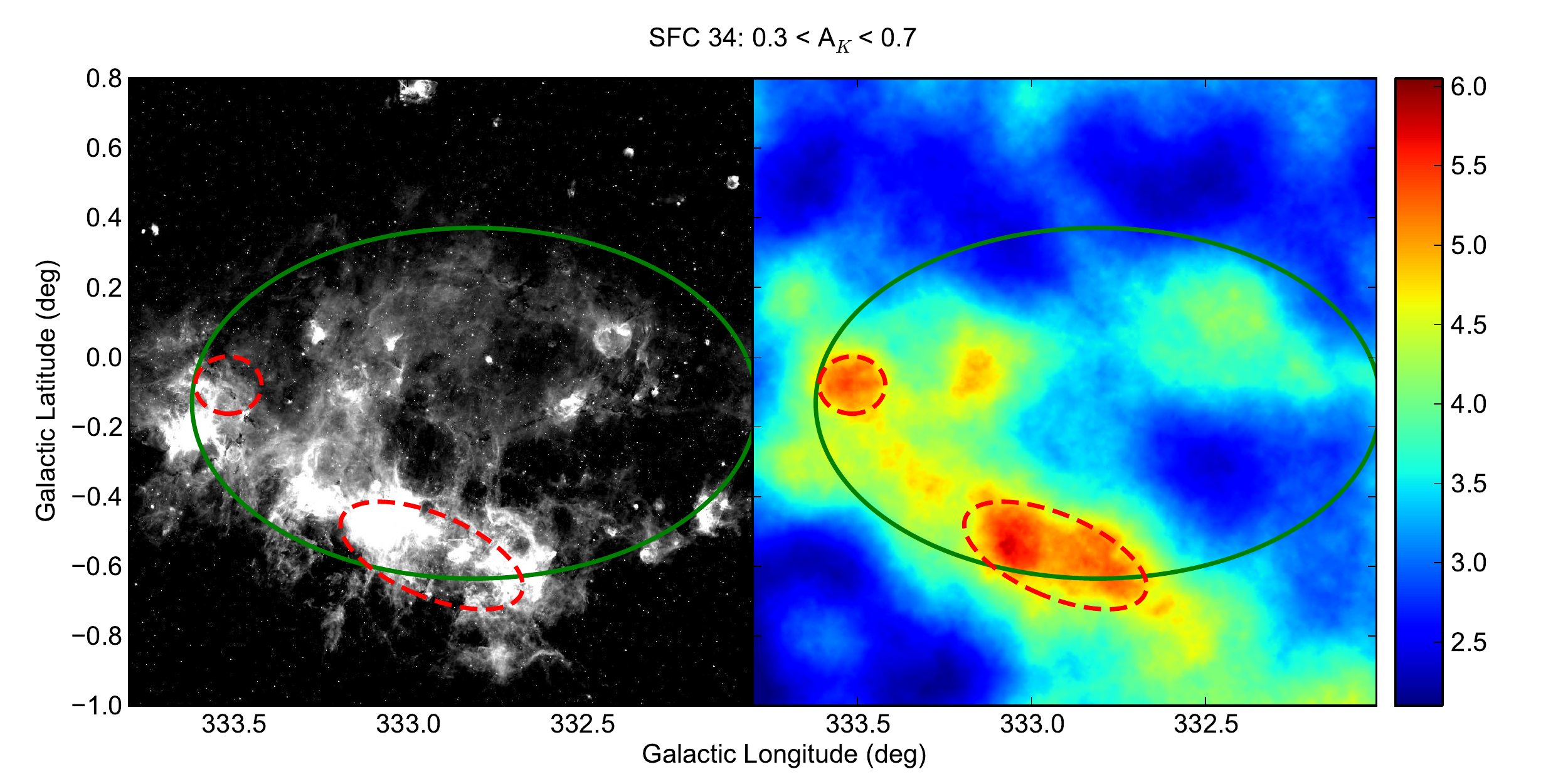}\\
    \includegraphics[scale=0.5]{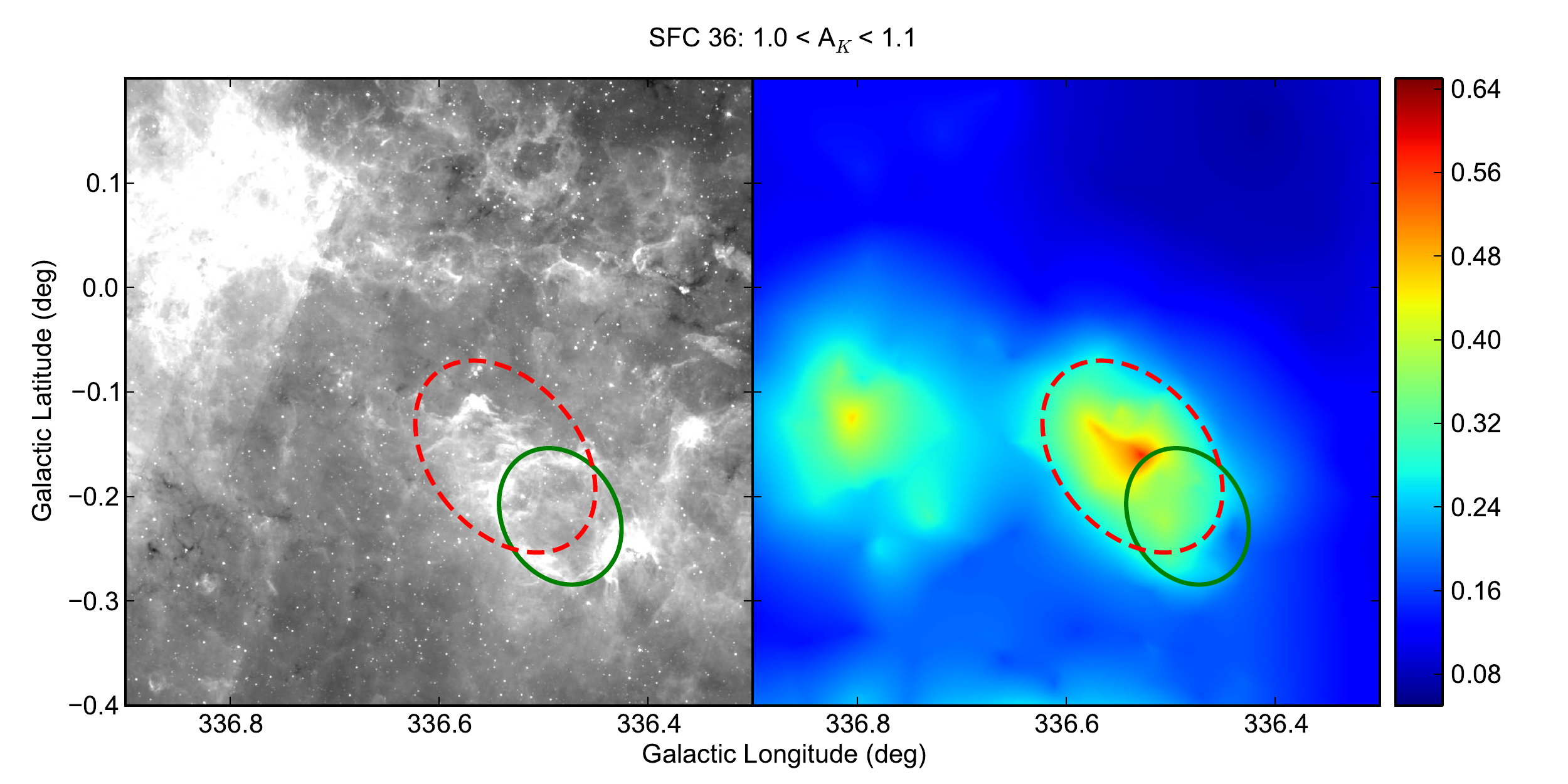}\\
  \end{center}
  \caption{Spitzer GLIMPSE 8$\micronm$ image (left) and the on-sky
    density diagrams (right) of the 2MASS point sources. The source
    and extinction ranges are indicated above each figure. The color
    bar and annotations are the same as Figure
    \ref{fig:sfc050607}. \label{fig:sfc333436}}
\end{figure*}

\begin{figure*}
  \begin{center}
    \includegraphics[scale=0.5]{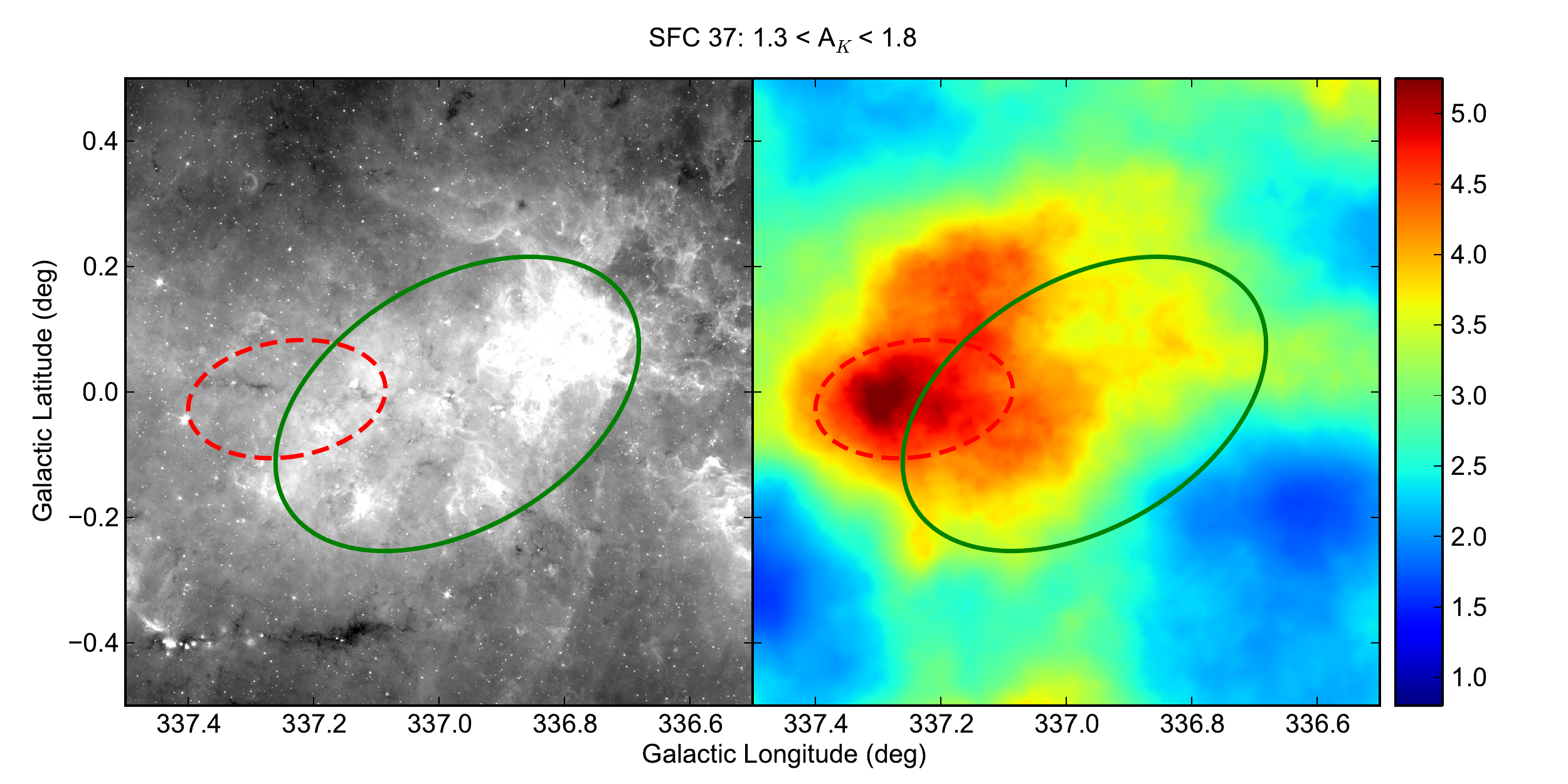}\\
    \includegraphics[scale=0.5]{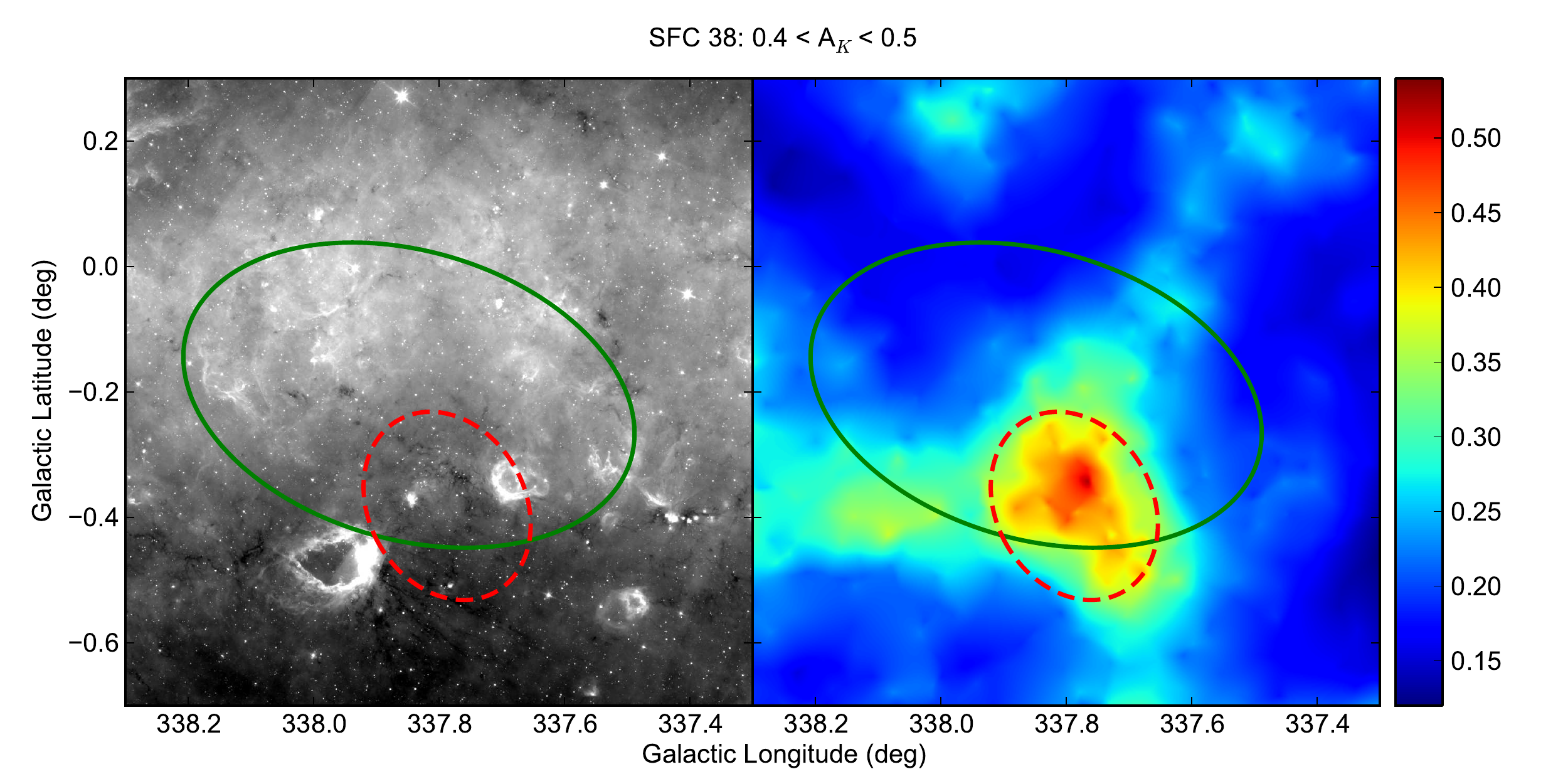}\\
    \includegraphics[scale=0.5]{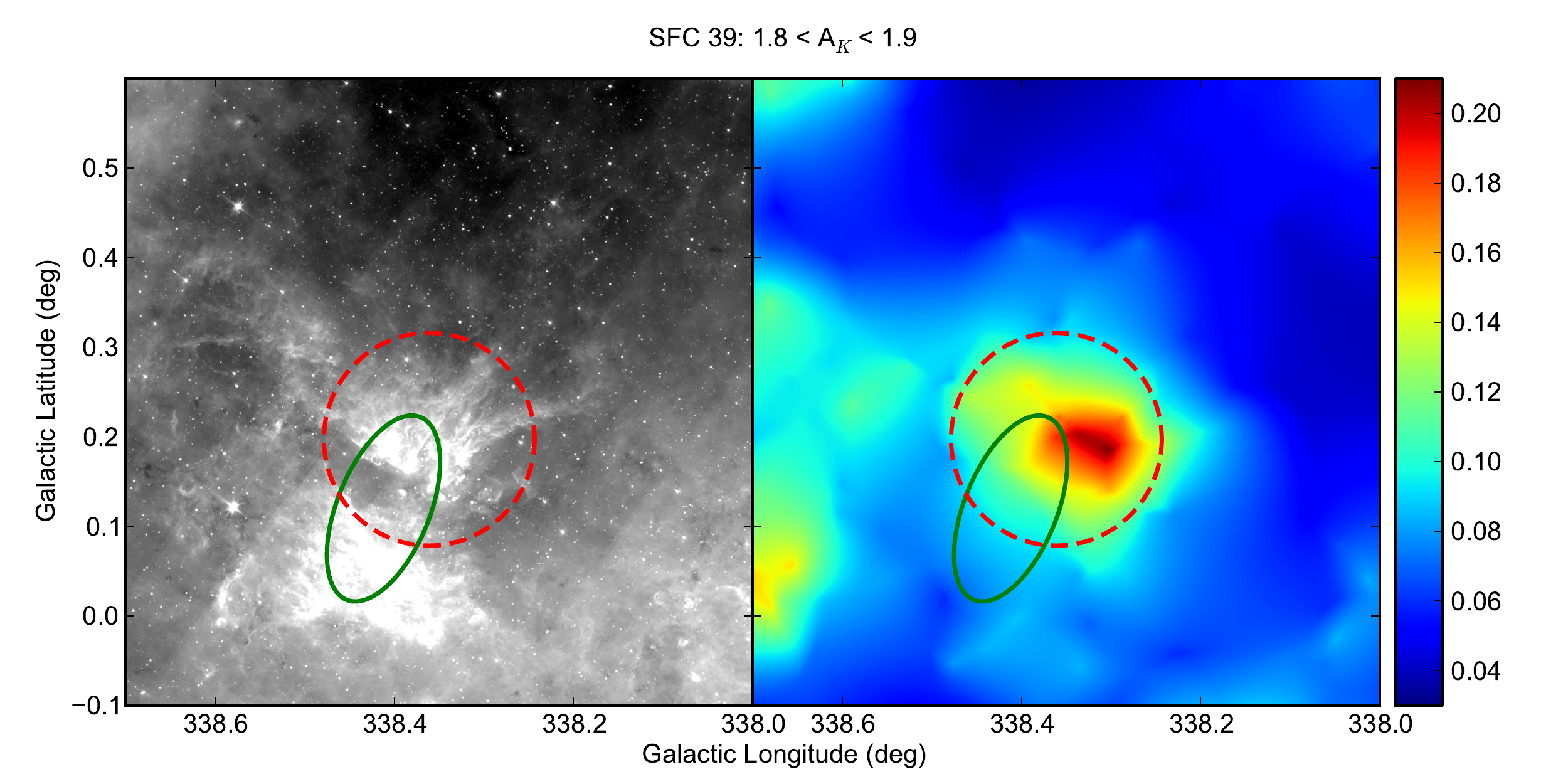}\\
  \end{center}
  \caption{Spitzer GLIMPSE 8$\micronm$ image (left) and the on-sky
    density diagrams (right) of the 2MASS point sources. The source
    and extinction ranges are indicated above each figure. The color
    bar and annotations are the same as Figure
    \ref{fig:sfc050607}. \label{fig:sfc373839}}
\end{figure*}

We present the stellar density diagrams and the 8$\micronm$ images
from the Spitzer GLIMPSE or Midcourse Space Experiment (MSX) in
Figures \ref{fig:sfc050607} to \ref{fig:sfc373839}. These images
indicate the SFC location from RM10 and the candidate cluster
location.

Many of the candidates are not centered on the nebulosity visible in
the 8$\micronm$ images. This may be an indication of the age of the
cluster; if the cluster is young, it would not have had sufficient
time to evacuate its surrounding material. Consequently, it would
appear at the same location as the 8$\micronm$ emission. If the
cluster is older, it would have evacuated its surrounding material and
likely blown a bubble. A cluster in this stage would appear separate
from the material it is illuminating. In addition, the SFCs identified
by RM10 are visually identified and may not correctly reflect the
location or the size of the SFC based on the limited data. If
confirmed, the candidate clusters will enable the refinement of the
SFC parameters.

\begin{figure*}
\begin{center} 
\includegraphics[scale=0.75]{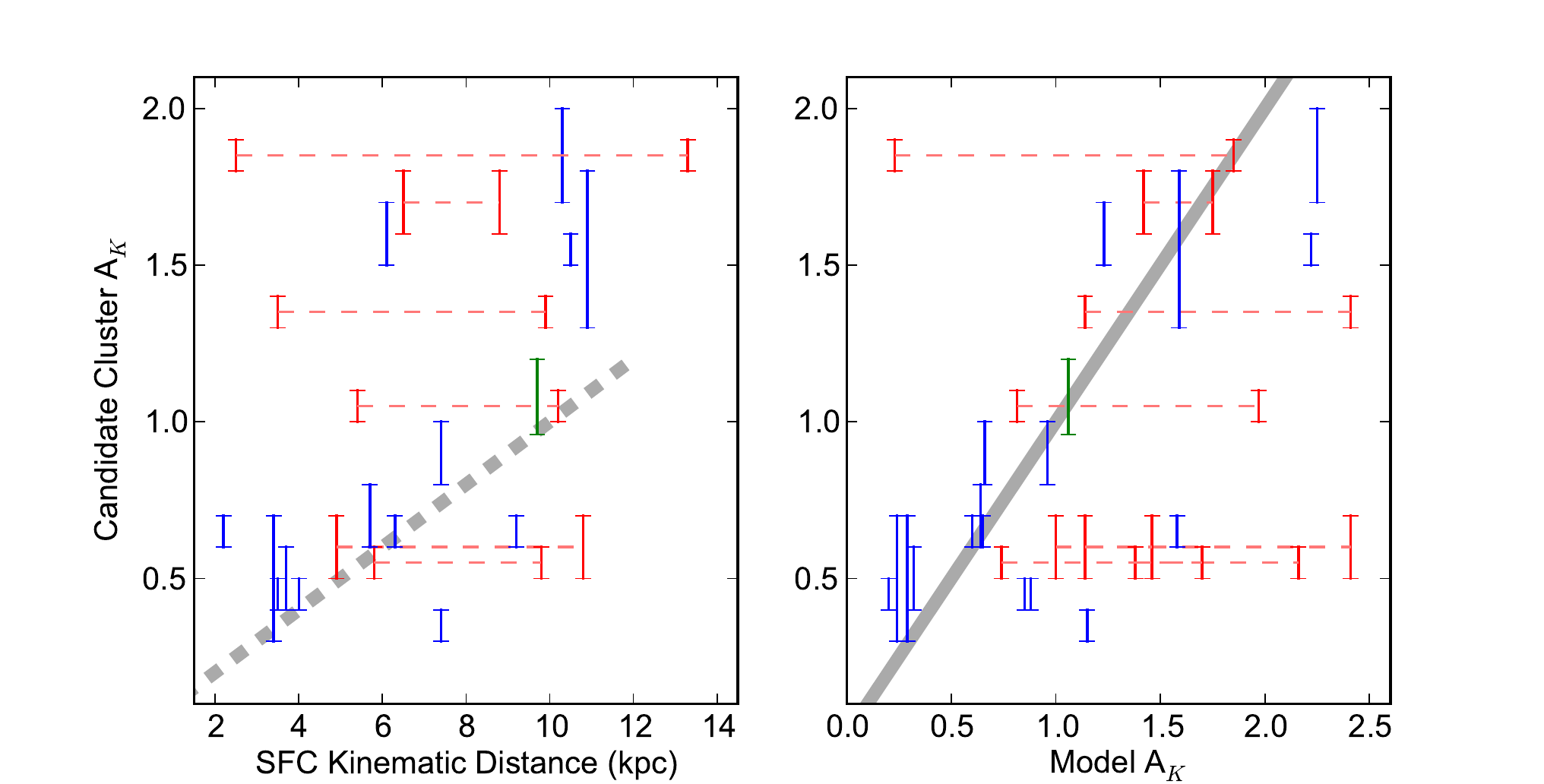}
\end{center}
\caption{A comparison between the extinction range of all identified
  candidate clusters to the kinematic distance of their host SFC from
  RM10 (left) and with the model extinction from
  \citet[][right]{marshall06}. The diagonal is indicated with the gray
  line on the right, and the local distance to extinction relationship
  ($A_{K}/D = 0.1$ mag kpc$^{-1}$) is indicated with the dashed gray
  line on the left. This relation breaks down for distances greater
  than 6 kpc, as compared to the relation between the candidate and
  model extinction that remains tight for all candidate distances.
  The connected red ranges indicate the SFCs without a kinematic
  distance ambiguity resolution. The green range indicates the
  location of the Dragonfish Association properties from
  \citet{rahman11a}. As no information about the kinematic distance
  was used in identifying the candidates, the correlation between the
  candidate and model extinction is not caused by a bias in the SSC
  method. \label{fig:distext}}
\end{figure*}

\begin{figure*}
\begin{center}
\includegraphics[scale=0.5]{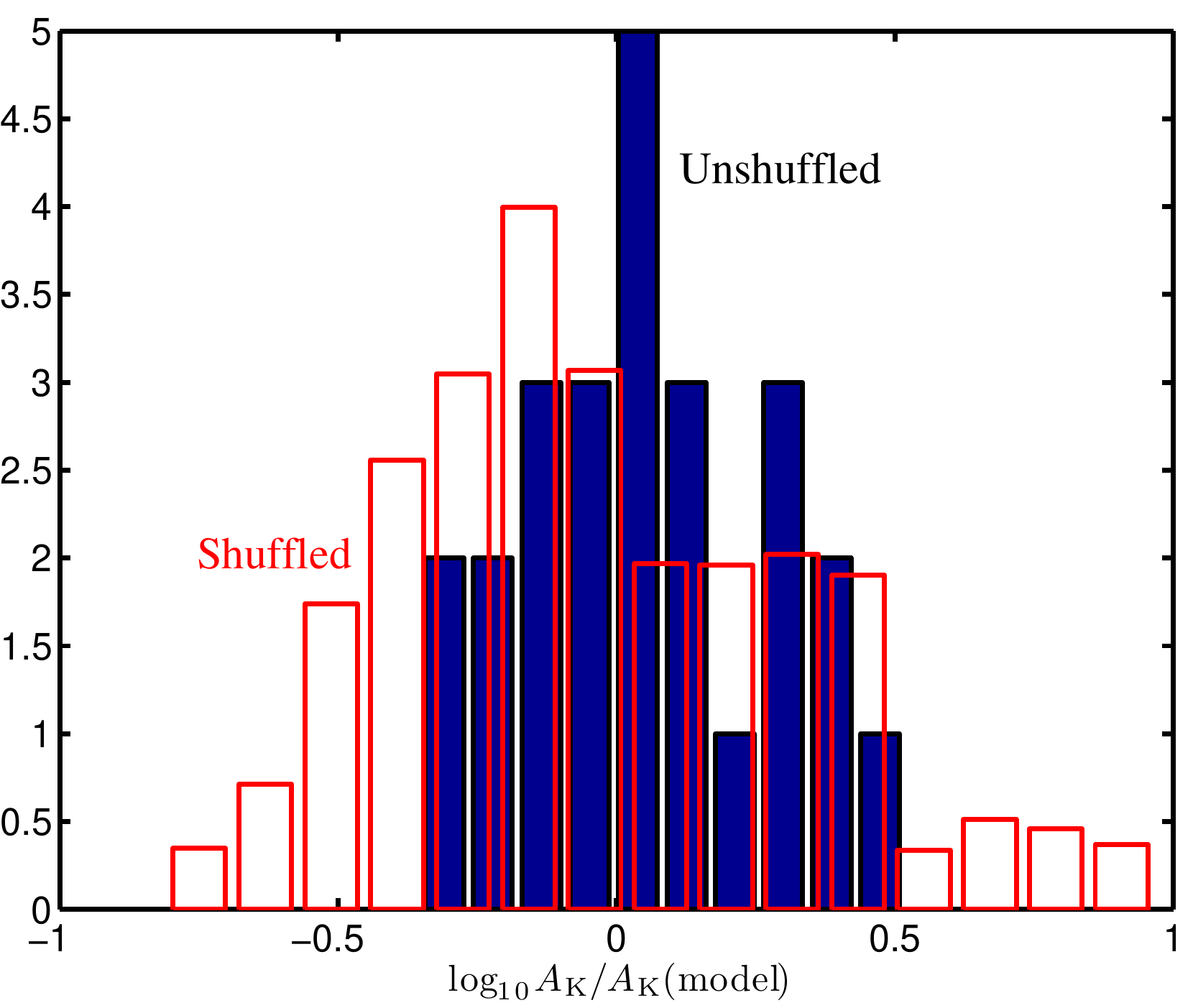}
\end{center}
\caption{A histogram of the ratios of the candidate cluster extinction
  to the modeled extinction from \citet{marshall06} in the filled
  histogram. The outlined histogram is the result of shuffling the
  candidate cluster extinctions 5000 times, normalized to the number
  of candidates in our sample. The larger scatter of the shuffled
  histogram provides additional support for the reality of the
  candidates. \label{fig:chrisfigure}}
\end{figure*}

Since the candidate extinctions were determined independently from the
position and distance to their host SFC, we can use the modeled
extinction from \citet{marshall06} as a consistency check. We present
the comparison of the candidate cluster extinction ranges with the
modeled extinction in Figure \ref{fig:distext}. To determine the model
extinctions, we use the SFC kinematic distance form RM10 in Table
\ref{tab:cclist}. We also indicate the possible model extinctions of
SFCs without a kinematic distance ambiguity solution and the location
of the Dragonfish association from \citet{rahman11a}. We find the
correlation coefficient of the two extinctions to be 0.75 when using
the more consistent distance in the cases of the distance ambiguity.
The correlation is not simply the result of choosing the most
consistent distance, however.  We have randomly shuffled the inferred
and modeled values of $A_{\rm K}$ five thousand times, choosing, for
each shuffling, the most consistent model extinction.  The results are
shown in Figure \ref{fig:chrisfigure}: without shuffling, the ratio of
modeled to observed extinction is 1.1$\pm$ 0.22 dex (standard
deviation in the logarithm); with shuffling, the ratio becomes 0.9 $\pm$
0.36 dex.  The increased scatter in the shuffled sample strongly
supports the validity of the candidate clusters, at least on a
statistical basis. This correlation provides strong evidence for the
reality of the candidate clusters. For sources with a distance
ambiguity, the modeled extinction provides support in favour of one of
the two possibilities. We find this correlation to be strikingly
better than the correlations between extinction and distance.

%\clearpage

%%% Stacked Figures 
%\include{stackedfigs-emapj}

%%% SFC 05, 06, 07

%%% SFC 08, 10, 12

%%% SFC 17, 19, 22

%%% SFC 23, 25, 26

%%% SFC 28, 30, 31

%%% SFC 33, 34, 36

%%% SFC 37, 38, 39

%%% Distance to Extinction Plot 

%%% Chris' correlation coefficient figure

\subsection{Previously Identified Clusters}
\label{subsect:previd}
An important test of the success and the sensitivity of the SSC method
is identifying known massive clusters and associations within
2MASS. For this purpose, we use the cases of NGC 3603, the powering
cluster of SFC 28, and Cygnus OB2. We discuss these two cases below.

\subsubsection{NGC 3603}
\label{sect:ngc3603}

\begin{figure*}
\begin{center}
\includegraphics[scale=0.75]{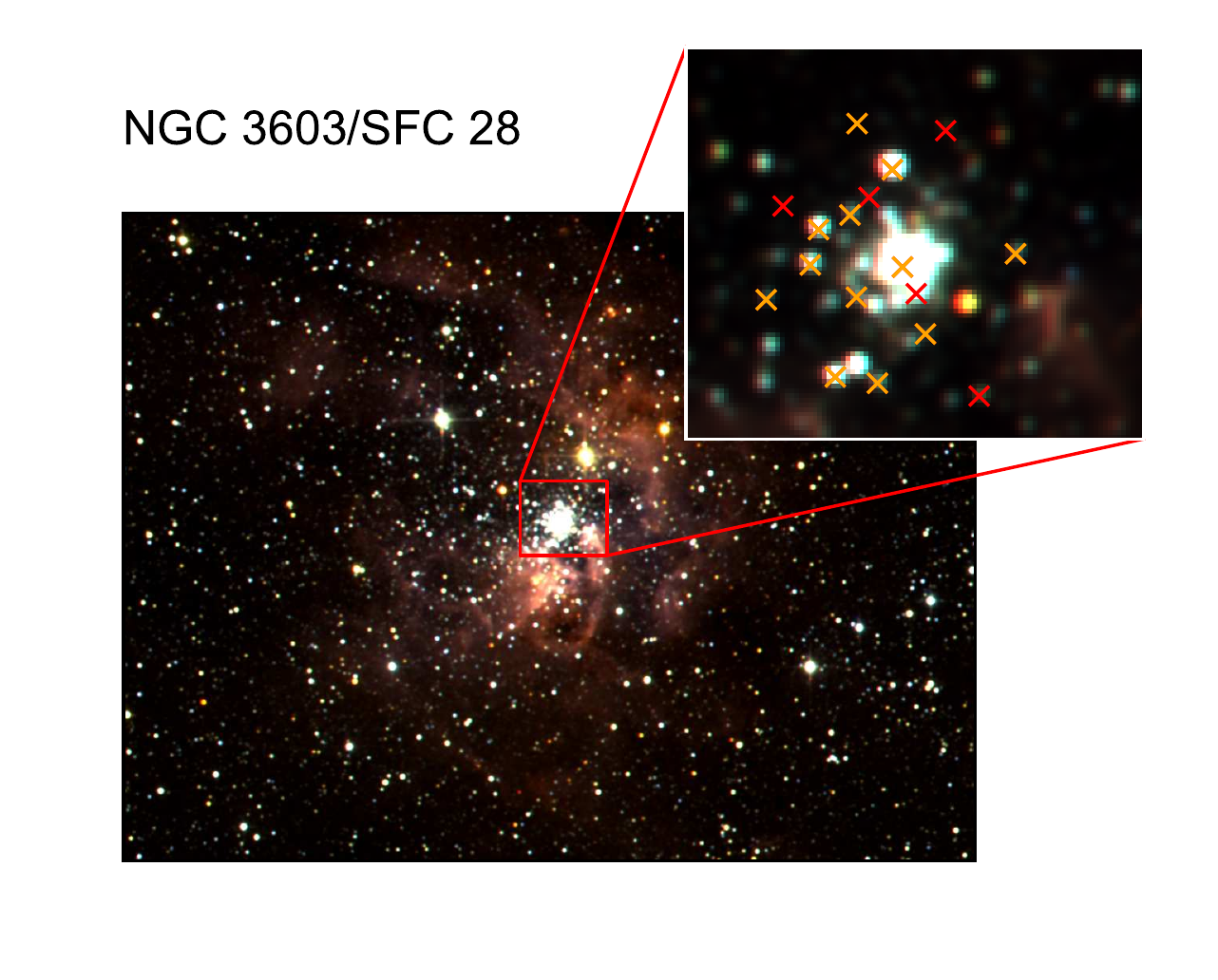}
\end{center}
\caption{NGC 3603 region from 2MASS in J (blue), H (green) and K (red)
  with the central cluster magnified (inset). The crosses indicate the
  locations of the only 2MASS point sources that pass the required
  quality cut within 33\arcsec of the cluster center. The orange
  crosses are the point sources that meet the color cut of the
  cluster, while those in red are the sources that do
  not. \label{fig:ngc3603}}
\end{figure*}

A notable cluster that was not detected by the SSC method is the
massive cluster within NGC 3603 that powers SFC 28. This very compact
cluster is mostly contained within 33\arcsec, requiring high spatial
resolution to discern individual members \citep{stolte04}. At the
resolution of 2MASS, the central cluster is highly confused, as can be
seen in Figure \ref{fig:ngc3603}. This prevents individual sources
from being identified, so they are missing from the 2MASS PSC. In
fact, only 17 point sources are identified in the central region that
pass the desired quality cut for the SSC method (inset of figure
\ref{fig:ngc3603}). Further filtering point sources to those with
colors corresponding to O-type stars at the extinction of NGC 3603, we
are restricted to 12 point sources. Although these 12 sources do
represent a local maximum in the color-selected source density within
the cluster's boundary, this is neither significant when compared with
the background variation, nor representative of the cluster's true
stellar population. In redder color cuts, this region is seen as a
void in point source density. This is consistent with our expectation
that the SSC method is insensitive to compact clusters, such as those
that can be easily identified visually.

Despite this, the SSC method is useful to investigate the entire area
of the NGC3603 natal cloud for a diffuse population surrounding the
dense cluster. We suspect that these are the candidate clusters we
identify in SFC 28, on the edge of the region. They appear at a
significantly higher extinction than NGC 3603 ($\Delta$A$_{K} \sim
0.4$). If they are not unrelated regions farther across the Galactic
plane, they could be regions deeply embedded within the cloud,
consistent with the CO morphology from \citet{grabelsky88}. In either
case, these candidates are not the powering source of SFC 28.

\subsubsection{Cygnus OB2}
\label{sect:cygob2}

To ensure that the SSC method can identify known OB associations, we
use Cygnus OB2 as a test case. This association, located in the outer
Galaxy, has a substantially reduced background stellar population in
comparison to the clusters in the inner galaxy. Therefore, the
association can be identified using total stellar density without
extinction cuts, as has been done by \citet{knod00}.

\begin{figure*}
\begin{center}
\includegraphics[scale=0.5]{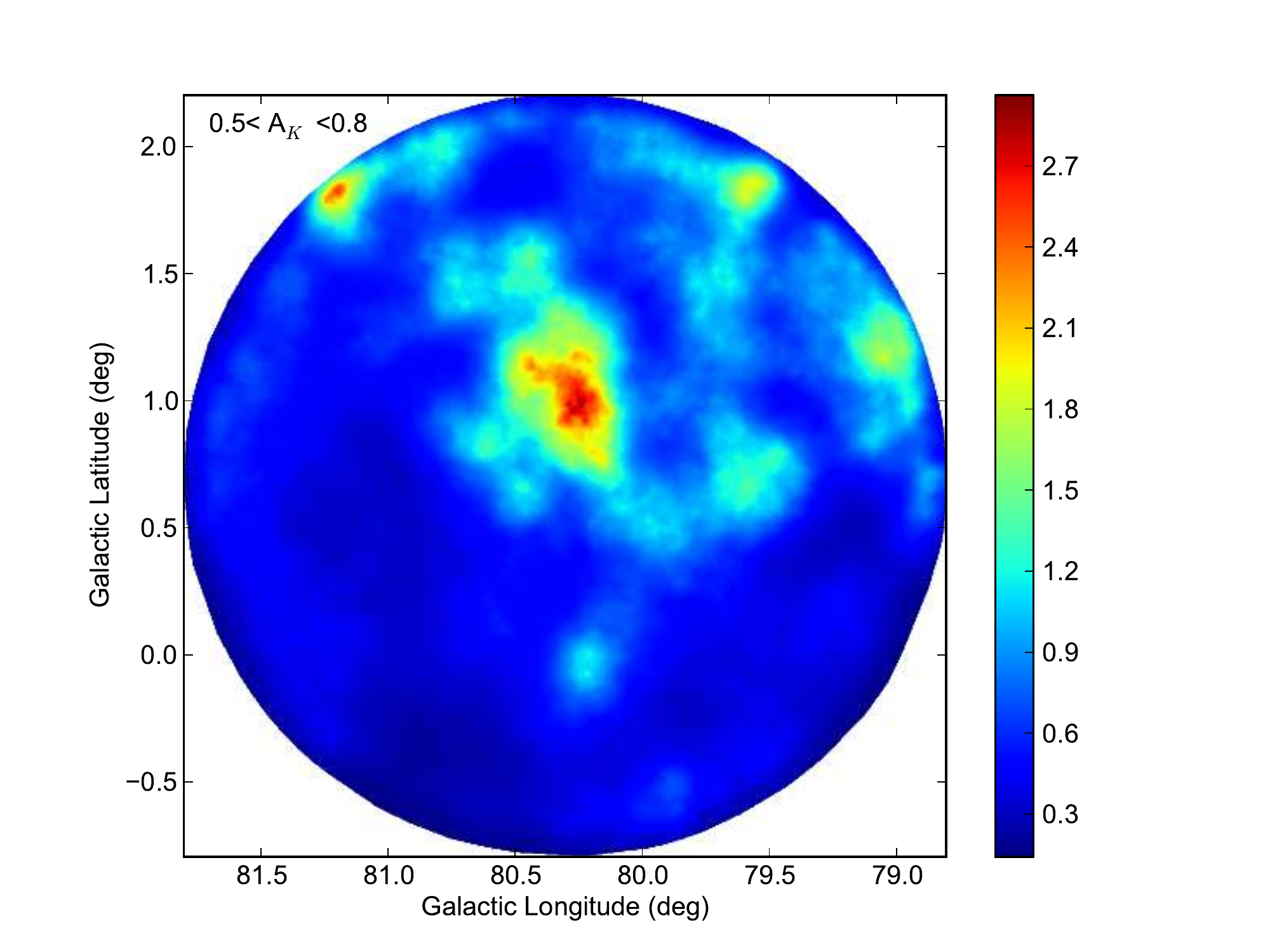}
\end{center}
\caption{The on-sky density diagram of 2MASS point sources between
  $0.5<$A$_{K}<0.8$ in the Cygnus OB2 region. The color bar indicates
  the point source density in sources
  arcminute$^{-2}$ \label{fig:cygob2}}
\end{figure*}

Using our method, we easily identify Cygnus OB2 (Figure
\ref{fig:cygob2}) at an extinction range of $0.5<$A$_{K}<0.8$, with
position, structure and extinction consistent with the values
determined by \citet{knod00}. The typical background level in the
field is 0.3--0.6 sources arcminute$^{-2}$, approximately
10\%--20\% of the peak density of the association. If we choose the
boundary of the cluster to be the isodensity contour at 1.2 per
arcminute$^{-2}$, the average rate of line-of-sight contaminating
sources within the association boundaries is $<0.28$, allowing for
high-yield extraction of association members from the field.

%%% Cyg OB2 

\subsection{Statistical Significance}

In \S\ref{subsect:previd}, we have discussed the success and
sensitivity of the SSC method using previously identified clusters and
associations. However, the problem that has plagued previous cluster
candidate identification methods has been the overwhelming number of
false positives. We conduct a test to determine the rate of false
positive candidate identifications through the SSC method.

We generate a list of 40 artificial SFCs with positions, dimensions
and position angles randomly chosen to match the distribution of
actual SFCs on the sky: the artificial locations are generated from a
uniform random distribution between $270^\circ < l < 90^\circ$ and $
-1^\circ < b < 1^\circ$. The position angles, semimajor axes and the
axial ratios were generated by producing linearly interpolated
cumulative distribution functions from the real SFC list.  Using the
SSC method, we attempt to identify ``candidate clusters'' for these
artificial SFCs, with the same criteria as in \S
\ref{sect:candid}. This test was conducted five times.  From the five
lists of artificial SFCs, the rate of identified candidates is $3.0\pm
0.6$ sources, which is a much smaller yield than the 22 actual SFCs
with identified candidates. It should be noted that the artificial
candidates may be actual clusters or associations related to less
luminous WMAP sources on the sky from \citet{murray10}.

In addition to the false positive test, the correlation between the
modelled extinction and the candidate extinction
(\S\ref{sect:candclust}; Figure \ref{fig:distext}) provides strong
support for the reality of the clusters identified by the SSC
method. Finally, the spectroscopic confirmation of the Dragonfish
Association \citep{rahman11b} provides additional, circumstantial
evidence of the clusters identified by this method. These independent
lines of evidence indicate that most ($\sim 86\%$) of our cluster
candidates are indeed associated with their host SFCs, and are their
dominant engines of ionization and dynamical power.

\subsection{Testing Cluster Extraction}
\label{subsect:exttest}

We conduct a pair of tests to determine any bias in the extracted
parameters of the candidate clusters. First, we test the accuracy of
the number of extracted sources as a function of extraction radius. We
select a 2MASS field centered on ($l$, $b$) = ($343.11\degr$,
$-0.50\degr$) with the extinction-based color cut $0.6 < A_{K}^\prime
< 0.7$ and insert randomly generated clusters. The background level of
sources in this field is 0.24 $\pm$ 0.11 arcminute$^{-2}$. The
clusters were generated with 40 visible stars in a 7\arcmin{} radius,
corresponding to a ``true'' extraction significance of $\sim$3 for the
entire radius of the cluster in this field and a central peak point
source density of 0.8 sources arcminute$^{-2}$. The worst case
scenario for a poor extraction yield is a shallow density profile;
this leads to the highest number of stars at large radii that may be
missed with the selection of a smaller extraction radius. To test this
worst case scenario, we model the cluster with a uniform stellar
density in three dimensions. We insert 1000 randomly generated
clusters into different random locations in the field. We extract each
of the clusters with extraction radii ranging from 0.5\arcmin{} to
20\arcmin{}.

\begin{figure*}
\begin{center} 
\includegraphics[scale=0.75]{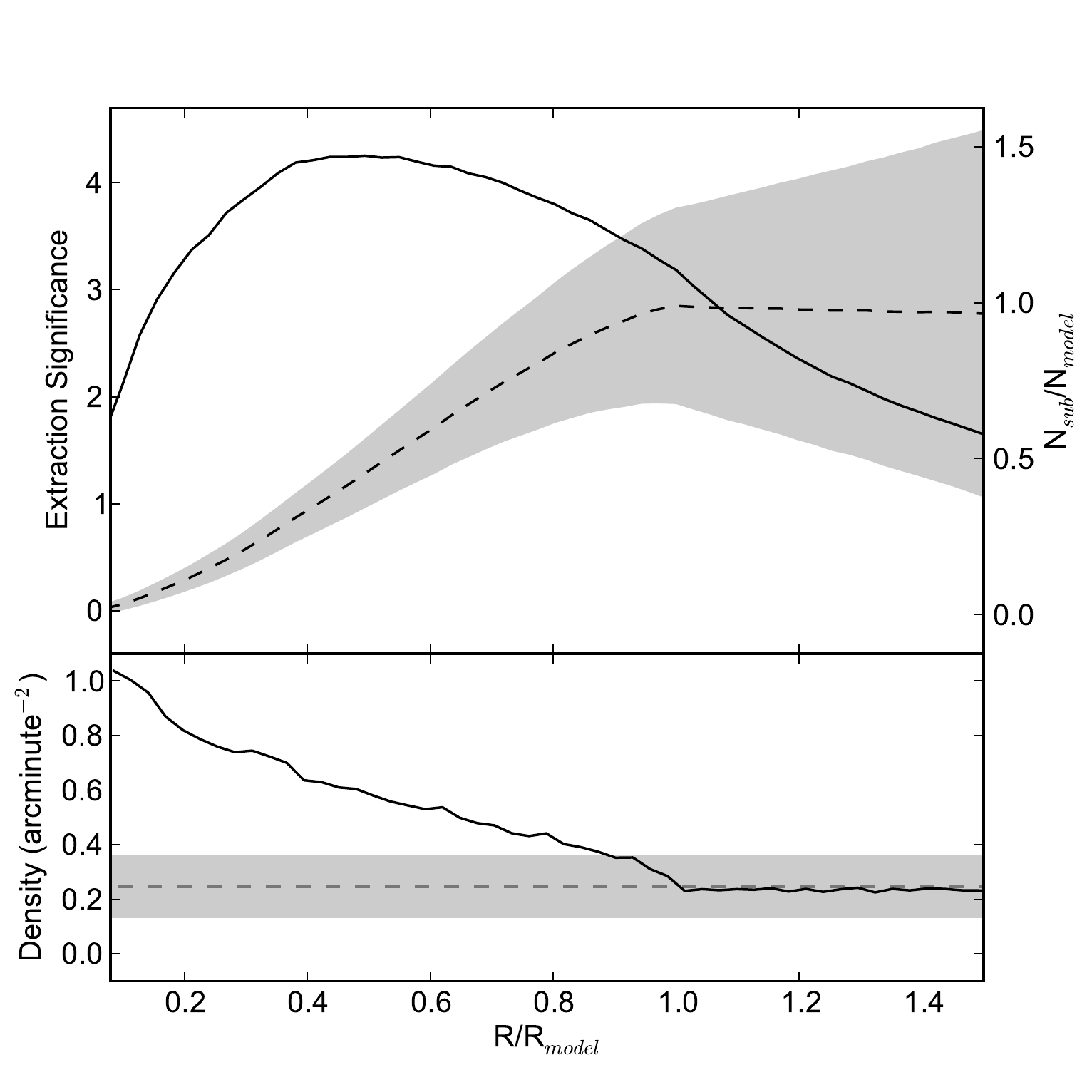}
\end{center}
\caption{The results of the model cluster extraction simulation. The
  mean extraction significance (solid line) and mean fraction of model
  stars extracted (dashed line) from all 1000 insertions are plotted
  in the top frame. The shaded area indicates the error in the
  fraction extracted based on the background variation. The mean point
  source surface density of all insertions is plotted in the bottom
  frame (solid line), with the background density level (0.24
  arcminute$^{-2}$; dashed line) and 1--$\sigma$ background variation
  (0.11 arcminute$^{-2}$; shaded region) indicated. The extraction
  radius (R) is normalized to the radius of the inserted modeled
  cluster (R$_{model}$ = 7\arcmin). \label{fig:testextraction}}
\end{figure*}

In Figure \ref{fig:testextraction}, we present the results of this
simulation. The 1--$\sigma$ extraction boundary corresponds to an
extraction radius that is $\sim$0.8 times the size of the model
cluster radius, and extracted cluster source numbers that are
$\sim$80\% of the inserted numbers. Arbitrarily larger extraction
radii produce extracted cluster source numbers that approach 100\% of
the inserted numbers, but with growing error at larger radii. When a
more conservative 2--$\sigma$ or 3--$\sigma$ extraction radius is
chosen, the extracted star numbers are 50\% and 30\% of the inserted
cluster numbers, underestimating the total cluster population given
this worst-case radial density model. This indicates that the chosen
selection boundaries are either correctly estimating or
underestimating the number of stars extracted.

Next, we conduct a test of the visual cluster identification and
extraction method to determine what biases exist in the selection of
the candidate clusters. For this test, we generate artificial clusters
with a random distribution of positions, number of sources, and
density profiles, place them in a real 2MASS field, and extract their
positions, sizes, and number of sources with background estimation in
the same manner as with the actual candidate clusters.

We generate 200 artificial clusters containing between 29 and 67
sources. We fix the truncation radius of the clusters at 7\arcmin{},
the median of the semimajor axes from the candidate clusters. We vary
the 2D-density profile of clusters using a S\'{e}rsic probability
distribution of the form:
\begin{equation}
P(R) \propto  \exp(-2 (R/R_{T})^{1/n_s})
\end{equation}
where $R_{T}$ is the truncation radius, and $n_s$ is the S\'{e}rsic
index. Larger S\'{e}rsic indices correspond to more centrally
concentrated distributions, while smaller indices produce more shallow
profiles. For simplicity in comparison, we choose discrete integer
values between 1 and 4 for the S\'{e}rsic indices.

We choose the same 3\degr{} 2MASS field ($l$, $b$ = $343.11\degr$,
$-0.50\degr$), and extinction-based color cut ($0.6 < A_{K}^\prime <
0.7$) from the earlier test. This field is free of identified
candidate clusters, but does contain the stratified Galactic structure
feature similar to that shown in Figure \ref{fig:galstruct}. We
represent the number of artificial cluster sources inserted as
$N_{input}$, and the number of field stars at that location as
$N_{bg}$. The number of sources estimated in the cluster after
background subtraction using the visual extraction method is
represented as $N_{sub}$.

Of the 200 artificial clusters inserted, 160 were successfully
identified and extracted. The centroiding of the identified clusters
is quite accurate; the mean difference between the inserted and
extracted central positions of the identified test clusters is
1.4$\pm$0.9 arcminutes and only 7 identified clusters have positional
differences greater than 3\arcmin. The 40 unidentified test clusters
are typically confused with Galactic structure and the size
distribution is skewed towards clusters with smaller numbers of
points. In addition, the unidentified test clusters are skewed towards
larger S\'{e}rsic indices. This indicates that the visual selection
method is biased against smaller clusters and those with peakier
density profiles. This is consistent with the expectation that we are
not sensitive to particularly compact clusters.

\begin{figure*}
\begin{center}
\includegraphics[scale=0.8]{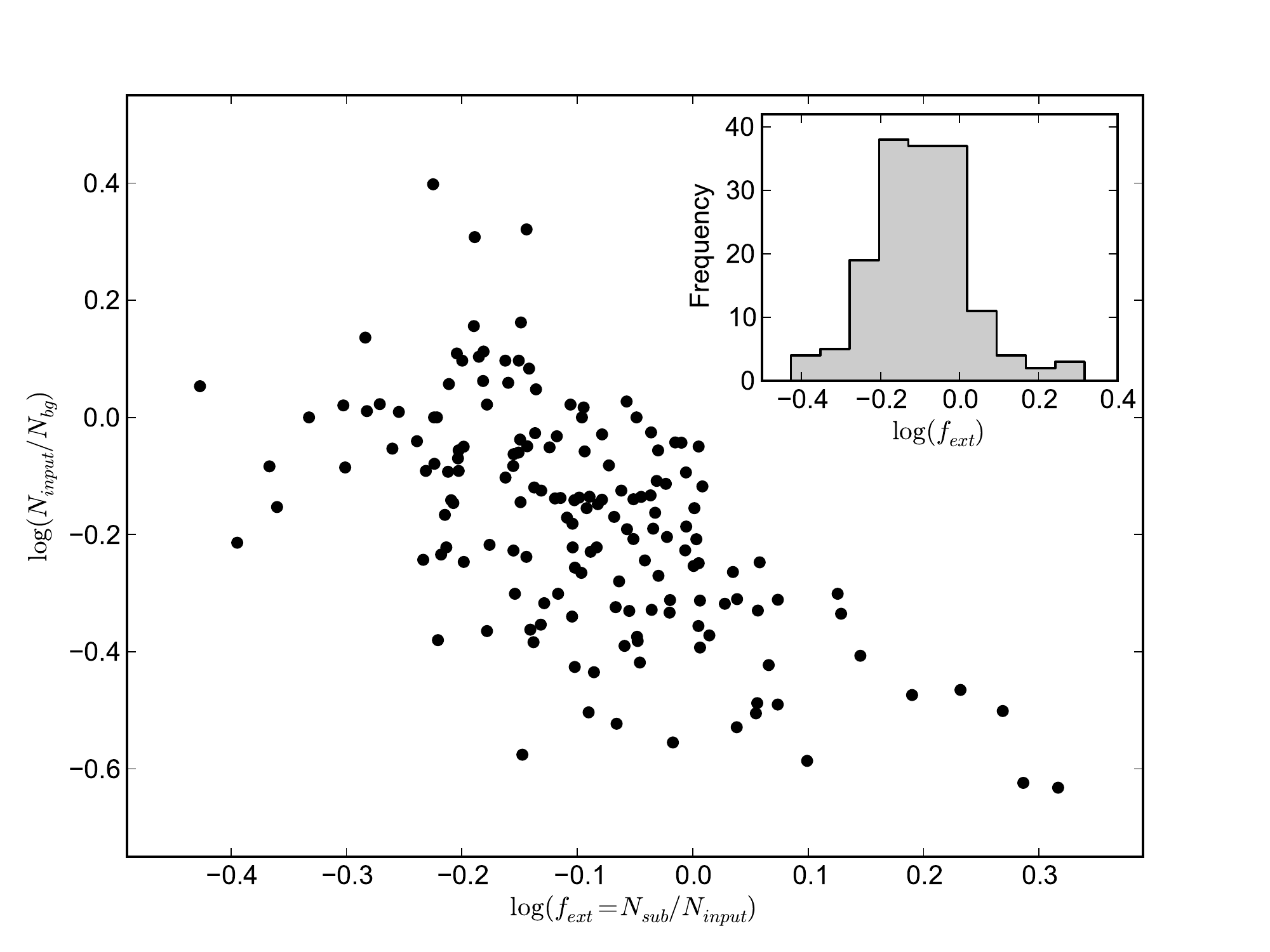}
\end{center}
\caption{The results of the cluster extraction test using artificial
  clusters in a Galactic 2MASS field centered on ($l,\, b$) =
  (343.11\degr, -0.5\degr). The ratio of the extracted point sources
  to the artificially inserted point sources, $f_{ext}$, vs. the ratio
  of the inserted point sources to the number of field sources in the
  background. We present the histogram of $f_{ext}$ of test clusters
  in the inset. The visual extraction method generally estimates a
  cluster population of between 60\% to 100\% of the inserted cluster
  population. The method only estimates a larger number of points than
  inserted into the cluster when the cluster is small in comparison to
  the number of background sources. \label{fig:insertclustertest}}
\end{figure*}

From the 160 artificial clusters extracted, we can determine the
accuracy of retrieving the initial cluster population. We find that
70\% of clusters have a fraction of extracted sources to the inserted
sources, $f_{ext} = N_{sub}/N_{input}$, between 0.6 and 1.0 with a
mean value of 0.84 (Figure \ref{fig:insertclustertest} inset). The
typical measured ES values of the artificial candidates range from 2.5
to 5.5. We find no strong correlations between $n_s$, $N_{input}$ or
$N_{bg}$ with $f_{ext}$. However, we find that there is an inverse
correlation of the ratio of $N_{input}$ and $N_{bg}$ with $f_{ext}$
(Figure \ref{fig:insertclustertest}); in the small number of cases
where we overestimate the size of the cluster (by up to a factor of
two), these clusters are a small fraction of the background level of
the region ($N_{input}/N_{bg} < 0.35$). Overall, these tests indicate
that the method is robust in recovering the original cluster
population and position, with a tendency to underestimate the total
population size.

%%% Test Figure 

\section{Mass \& Luminosity Estimates}
\label{sect:phys}

We estimate the physical properties of the candidate clusters to
assess if the identified candidates can produce the observed free-free
flux of their parent WMAP source (Table \ref{tab:physprop}). We note
that the stated boundaries from the candidate identification and
extraction method are only visually identified based on a clear
overdensity from the local background variation
(\S\ref{sect:candclust}).  Consequently, the extracted clusters will
likely exclude the diffuse outskirts of the clusters or associations,
especially when the background level of contaminating sources is
particularly high. Such diffuse populations can produce significant
contributions to the total complex population, as has been recently
observed in the Carina Nebula where previously undiscovered OB stars
were identified in X-ray observations, doubling the massive stellar
population \citep{povich11}. Although we found in the previous section
that our method is likely to underestimate clusters' true population,
we adopt a conservative approach and do not attempt to correct for the
extraction efficiency.

We estimate the least massive stars that are extracted in the
candidate clusters by using the observed 2MASS \kmag-band magnitude
limit from \S \ref{subsect:2mass}, with the mean extinction range of
the candidate and the kinematic distance of the SFC from RM10. In
cases where the kinematic distance is ambiguous, we use the distance
most consistent with the Marshall extinction (see Figure
\ref{fig:distext}). We use the absolute magnitude tabulations from
\citet{martins06} for O-type stars and from \citet{pickles98} for all
other stars. We also use \citet{martins05} for the O-type star mass
determinations and \citet{andersen91} for all other stars. From the
mass of the least massive star and the number of stars extracted in
the candidate cluster ($N_{tot} - N_{BG}$ in Table \ref{tab:cclist}),
we determine a total cluster mass ($M_{*}$) and ionizing luminosity
($Q_{0}$) integrating over an IMF. To maintain consistency with
previous work, we use the modified version of the \citet{muench02} IMF
from \citet{murray10}. The mass of the candidate was estimated by
scaling the Muench IMF to the number of stars down to the least
massive star extracted, and then integrating over the mass function to
determine the total mass. The ionizing luminosity was determined by
using the mass-to-ionizing luminosity relationship from
\citet{murray10}. Note that we do not make any correction for the
multiplicity of sources.

Finally, we determine the expected free-free flux contribution of the
candidate at $\nu = 90$ GHz. We also visually classify the candidates
based on their location with respect to the identified SFC and the 8
$\micronm$ morphology. We indicate that they are either
``centrally-located'' if they appear towards the center of a
shell-like structure or the entire SFC, or ``shell-located'' if they
appear on the edge of a shell or SFC.

In Table \ref{tab:cclist}, $N_{sub}$ is the number of sources in the
candidate cluster once the background level is subtracted
($N_{TOT}-N_{BG}$).  We include the Dragonfish Association properties
from \citet{rahman11a}. For the WMAP sources with multiple identified
candidate clusters, we include totals for the cluster masses, ionizing
luminosities and free-free emission fluxes.

We note that the latest stellar type of all extracted sources is
between O- to A-type stars. All of these hot stars will have similar
Rayleigh-Jeans colors in the NIR. Therefore, we do not expect cluster
members to be excluded based on having colors that fall outside of the
color cut.

The total cluster masses of the candidates range from $150$ to $10^5$
M$_{\sun}$; perhaps not surprisingly, some appear to be low-mass
objects unrelated to their host regions. In these cases, the estimated
ionizing luminosity and free-free emission flux are overestimated as
the IMF is poorly sampled on the upper end and the mass-to-ionizing
luminosity conversion breaks down. This has been discussed in detail
by \citet{kennicutt89}. Therefore the free-free flux contribution from
these clusters will be much smaller than those tabulated since the
mass-to-ionizing luminosity relation is unsuitable for a poorly
sampled IMF. By ionizing luminosities of log Q$_{0} \sim 50$ or
cluster masses greater than 1600 M$_{\sun}$, $>10$ O-type stars are
required to ionize the complex and the IMF should be sufficiently well
sampled to allow the accurate use of the mass-to-ionizing luminosity
relationship \citep{martins06}.

We find the sum of the estimated free-free emission fluxes to be
roughly consistent with the WMAP measurement for G34, G37, G298, G327,
and G337. For G30 and G49, the identified candidates appear to be
centrally located, but the smaller-than-expected mass, ionizing
luminosity, and free-free flux most likely come from a diffuse
component of the cluster/association. Should these candidates be
confirmed, an in-depth investigation, possibly in the X-rays, will be
required to identify the extended or diffuse population. For G283 and
G327, only shell-located regions are identified. In the case of G327,
the lack of sharp shell structure in the bubble in 8 $\micronm$ and
2MASS stellar overdensity structure (see Figure \ref{fig:sfc333436},
top panel) may indicate that the original driving cluster of the
bubble has evolved beyond its ionizing luminous stage, and the
ionizing population of the complex is a more diffuse, triggered
population.  In source G291, we identify two possible triggered
clusters; however the dominant cluster in the region, NGC3603, has
been previously identified and was not detected by the SSC method (see
\S \ref{sect:ngc3603}).

In some cases, such as G332, multiple candidates are identified at the
same (or reasonably close) distances and extinction ranges. These
candidates are likely associated and may represent more widespread
star formation across a molecular cloud or complex. Accordingly, the
total embedded cluster mass in these situations would be the sum of
the individual ``clustered'' regions, plus the diffuse component that
we do not recover from the extraction process.

The total expected free-free emission luminosities of the candidates
in G24 and G332 substantially exceed the measured flux from
WMAP. There are at least six possibilities to account for this: (1)
one or more of the candidates within the WMAP source are false
positives, (2) the extraction has overestimated the number of member
stars by a substantial fraction, (3) the kinematic distance to the SFC
is closer than the actual distance, (4) the cluster is older than 3.9
Myr, the age at which the most massive stars have left the main
sequence \citep{bressan93}, (5) a higher-than-average fraction of the
ionizing photons are absorbed by dust, and (6) an unusually large
fraction of the ionizing photons escape the region, or escape the
Galaxy. To determine whether it is case (1) or (2), follow-up
astrometric or spectral confirmation of the cluster is required. In
case (3), the distances to the ionizing sources can be unambiguously
determined though trigonometric parallax, if strong maser sources
associated with the \ion{H}{2} regions are available.  In case (4),
the evolutionary state of the candidate requires accurate
determination of the spectral type of candidate sources to determine
the earliest stars in the cluster and their individual evolutionary
state. If this is the case, these clusters may harbor young
core-collapse remnants of very massive stars, such as black holes or
magnetars.

With these caveats, we find candidates to central powering clusters
that can account for the observed WMAP free-free emission flux within
G24, G30, G34, G37, G49, G298, G332, and G337. Of the original WMAP
sources investigated in RM10, only G10 does not have any candidate
clusters identified through this method, either centrally located or
on the shell.

\section{Completing the Upper End of the Galactic Cluster Mass Function}
\label{sect:gcmf}

Our knowledge of the upper end of the Galactic cluster mass function
(CMF) has long been recognized as incomplete. This has been
established through comparison to galaxies with similar masses and
star formation rates as the Milky Way \citep{larsen00, larsen09}, and
comparing the known cluster/association population with the inferred
\hii mass function \citep{mckee97, hanson08b}. These comparisons show
that up to 90\% of young, massive clusters and OB associations are yet
to be found.

Table \ref{tab:census} presents the current census of known massive,
young ($\la 4$ Myr) clusters and OB associations in the Galaxy. This
census includes all known clusters and associations with masses
similar to or greater than $10^{4}$ M$_{\sun}$, in addition to the
massive candidates from this work. The age limit of 4 Myrs ensures
that even the most massive stars, which produce the bulk of the
ionizing luminosity, are still on (or near) the main sequence
\citep{murray10}. This effect is prominently seen in the case of
Westerlund 1, where despite the large cluster mass ($> 10^{4.7}$
M$_{\sun}$) and rich population of Wolf-Rayet and late-O type stars
\citep{brandner08}, the cluster is not detected as a prominent
free-free emission source in WMAP \citep{murray10}. \ If the
candidates from this work are confirmed with the minimum masses
estimated, the census of young, massive clusters and OB associations
is doubled with 7 new candidates.

Notably, the age restriction of the census excludes the red supergiant
clusters recently discovered with ages $>10$ Myrs and masses $> 10^4$
M$_{\sun}$ \citep{figer06, davies07, clark09}. The massive stellar
component of these clusters has entirely evolved off the main
sequence, and consequently the clusters have an insignificant ionizing
luminosity. This is corroborated by the lack of any reprocessed
free-free emission towards the clusters from the WMAP free-free
emission maps \citep{murray10}.

\subsection{Estimates of Completeness}

The completeness of the census can be estimated through CMFs from
previous work for other galaxies in the visible wavebands, such as the
Antennae and M83 \citep{fall09, chandar10}. \citet{ivanov10} estimate
that the Milky Way contains $\ge 81 \pm 21$ clusters with masses of
$10^4 - 10^5$ M$_{\sun}$ and ages up to $\sim$25 Myr, assuming all
clusters within 6 kpc are known \citep{pisk08}. This estimate can be
scaled to our 4 Myr age limit using the age distribution from
\citet[][ $\frac{dN}{d\tau} \propto \tau^{-1}$]{chandar10} with a
minimum age of 1.3 Myr based on the youngest clusters observed. The
scaled estimate is $31$ clusters in the Milky Way, of which
\citet{larsen09} would predict that 8 are bound. This is consistent
with the estimate made by \citet{mckee97}, who estimate the number of
OB associations with M $> 10^{4}$ M$_{\sun}$ (or Q$_0 > 10^{50.8}$
s$^{-1}$) to be 29.

Another independent measure of completeness is a comparison to the
\hii region luminosity functions of other galaxies. We compare to the
\hii region luminosity functions from \citet{kennicutt89}, using a
minimum H$\alpha$ luminosity of $L(H\alpha) > 10^{38.6}$ erg s$^{-1}$
which is similar to the $10^4$ M$_{\sun}$ cutoff. In the most
analogous cases to the Milky Way in terms of morphological class, NGC
6384 (an Sb galaxy) contains 38 sufficiently luminous \hii regions,
while NGC 7741 (an SBc galaxy) contains 17. \citet{smith89} show that
the \hii luminosity function of the Milky Way is comparable to those
of Sb-c galaxies, indicating that using the Sb-c galaxies as a guide
is a valid estimation of the total young, massive clusters in the
Galaxy.

Both of these estimates of the total number of massive ($> 10^{4}$
M$_{\sun}$) clusters/associations in the Galaxy are comparable to 15,
the number of associations in our census, of which seven are presented
here (Table \ref{tab:census}). Since the estimates of the number of
young, massive clusters expected is between 17-38, we infer from this
that our knowledge of young, massive clusters in the Milky Way is now
about half complete.

\subsection{Remaining Galactic Plane Targets}

This work is based on the luminosity-selected sample of SFCs from RM10
on the basis that their ionizing luminosities are from young, massive
clusters hosted by the SFCs. However, this work was not a
comprehensive search through the entire Galactic plane. We can
determine the likely regions of young, massive clusters that remain
hidden by comparing our census to the catalog of giant \hii regions
from \citet{conti04}.  As noted in RM10, previously identified
individual regions closely packed together are likely associated in
larger complexes (see \S 2.1 in RM10 for a detailed discussion). For
the purpose of this comparison, we group the giant \hii regions from
\citet{conti04} into 27 degree-sized zones in the Galactic plane. Each
Galactic plane zone consists of 1--7 giant \hii regions.

Of the 27 Galactic plane zones, young massive ($\ga 10^4$ M$_{\sun}$)
clusters or associations have not been identified in 18. Within the 18
zones without young massive clusters, 6 have previously known powering
clusters with masses less than $10^4$ M$_{\sun}$, and 5 have
candidates identified in this work also less than $10^4$
M$_{\sun}$. We list the zones lacking a young, massive cluster in
Table \ref{tab:missing}, tabulating the Galactic Plane Region (in
Galactic Longitude), their kinematic distance(s), and their names
where applicable. For the 5 zones with candidates identified in this
work, they have lower mass limits $>10^{3.2}$ M$_{\sun}$. Because our
search is not sensitive to the diffuse outskirts of
clusters/associations, it is possible that the candidates in the 5
zones have masses larger than $10^4$ M$_{\sun}$. Further observations
and analysis are required to better constrain the total masses of
these clusters.

The advent of new Galactic plane NIR surveys, such as the UKIRT
Infrared Deep Sky Survey \citep[UKIDSS; ][]{lawrence07} and the Vista
Variables in the Via Lactea Survey \citep[VVV; ][]{minniti10}, will
enable more sensitive probes for clustering than has been possible
with 2MASS. Both UKIDSS and VVV are significantly deeper than 2MASS
($\sim$4 magnitudes in the K-band), with a much smaller point source
function ($\sim$4 times smaller). Consequently, applying the SSC
method in these surveys would enable the detection of more highly
extinguished and less massive clusters, since the greater sensitivity
allows for more highly reddened and intrinsically less luminous stars
to be detected. In addition, these surveys would better probe more
compact clusters, where the 2MASS point source function size has
rendered them confused. However, two challenges will arise in using
deeper surveys; because they are more sensitive to the less luminous
members of a cluster, the assumption of Rayleigh-Jeans NIR colors will
break down and a more complex search strategy following the expected
color of the extinguished main-sequence will be required. Secondly,
the increase in sensitivity will also increase the number of field
sources; to better separate field and cluster sources, the explicit
use of the luminosity information will be required. However, the
potential in these new surveys is not limited to finding the remaining
Galactic massive cluster and association population, but also better
characterizing the candidate clusters and associations detected in
this work.

We note that this work searches for candidate clusters with
extinctions of $A_{K} < 2.0$, while the extinction model from
\S\ref{sect:candclust} indicates that significant portions of the
Galactic plane will have extinctions of greater than that. This
includes a number of candidates from this work if you accept the far
kinematic distance from the SFC. The currently used 2MASS data is not
sufficiently populated with highly-reddened stars to allow the SSC
work due at high extinctions. However, the deeper NIR surveys will
allow searches for more highly extinguished clusters and associations.

Twelve of the 27 Galactic plane zones have not been probed in this
work, leaving a substantial discovery space for the SSC method to be
applied. Recently, \citet{lee12} have identified SFCs in the remaining
WMAP free-free sources not originally covered in RM10, which provides
a new sample of morphologically-identified regions to probe.
Investigating this remaining SFCs may enable the completion of the
entire upper end of the Galactic CMF.

\section{Conclusions}
\label{sect:conclusions}

We have used the 2MASS point-source catalog to search for massive
clusters and association candidates within the SFCs from WMAP, looking
for the brightest stars, taking into account the line of sight
extinction. Including the previous work of \citet{rahman11a}, we find
that 22 of the 40 SFCs host a candidate cluster that has been
identified through our method, either central to the SFC or on its
periphery. The extracted candidates have semimajor axes of 3\arcmin{}
-- 26\arcmin. The average extraction significance of the candidates is
2.4. The candidates with smaller extraction significances are selected
based primarily on morphology.

We discuss the case of two previously identified
clusters/associations: NGC 3603, which is unidentified through our
method due to the compact nature of the cluster, and Cygnus OB2, which
we readily identify and whose properties we accurately constrain. We
find the false positive rate to be $3.0 \pm 0.6 $ out of 22
identified. We test the visual identification method and find it to be
robust in determining the population of the cluster. The strong
correlation between the candidate extinctions and the modelled
extinction based on the SFC distances provides strong support for the
reality of the clusters. This is reinforced by the spectroscopic
confirmation of the Dragonfish Association, originally identified the
same method of \citet{rahman11b}.

We estimate the masses and luminosities of the candidates, showing
that the candidates can account for the observed WMAP free-free flux
for most of the investigated sources. Of the WMAP sources
investigated, only one (G10) lacks any candidate clusters or
associations. Finally, we provide a Galactic census of the young,
massive clusters and OB associations. With the newly identified
candidates from this work, we estimate that our knowledge of the
young, massive cluster population is about half complete.  We indicate
locations on the Galactic plane possibly hosting the remaining
unidentified massive clusters and associations.

Future astrometric, spectroscopic and/or X-ray follow-up observations
can confirm the candidates, measure their physical properties and
determine their environmental impact. With such confirmation, the
candidate clusters and associations within the most luminous SFCs in
the Galaxy will enable the more accurate determination of the upper
end of the Galactic CMF.

\acknowledgements 

We thank B. Arsenault and H. White for assistance in the preliminary
stages of this work. We also thank N. Murray, R. Kennicutt, J. Graham,
and P.G. Martin for the many helpful discussions and comments. This
publication makes use of data products from the Two Micron All Sky
Survey, which is a joint project of the University of Massachusetts
and the Infrared Processing and Analysis Center/California Institute
of Technology, funded by the National Aeronautics and Space
Administration and the National Science Foundation. C.D.M and
D-S.M. acknowledge support from the Natural Science and Engineering
Research Council of Canada. This paper was studied with the support of
the Ministry of Education Science and Technology (MEST) and the Korean
Federation of Science and Technology Societies (KOFST).

\bibliography{ccpaper}

%%% All Tables
\include{tab1-emapj}

\include{tab2-emapj}

\include{tab3-emapj}

\include{tab4-emapj}

\end{document}

%% file: tab1-emapj.tex
\begin{deluxetable*}{ccccccccccccccc}
\tablewidth{0pt} 
\tablenum{1}\label{tab:cclist}
\tabletypesize{\scriptsize} \tablecaption{Candidate
  Clusters/Associations}
\tablehead{\colhead{}   & \colhead{$l$}   & \colhead{$b$}                      & \colhead{D\tablenotemark{b}}     & \colhead{smaj}     & \colhead{smin}      & 
  \colhead{$PA$}        & \colhead{}      & \multicolumn{2}{c}{$A_{K}^\prime$} & \colhead{$A_{K}$\tablenotemark{c}} & \colhead{}         & 
\colhead{}              & \colhead{}      & \colhead{}   \\
\colhead{SFC}           & \colhead{(deg)} & \colhead{(deg)}                    & \colhead{(kpc)}                    & \colhead{(arcmin)} & \colhead{(arcmin)}  & 
\colhead{(deg)}         & \colhead{$j$}   & \colhead{Min}                      & \colhead{Max}                      & \colhead{(Model)}  & \colhead{$N_{TOT}$} & \colhead{$N_{BG}$} & 
\colhead{$\sigma_{BG}$} & 
\colhead{ES}}  
\startdata
5a                      & 23.108          & -0.439   & 4.9, (10.8)    & 6    & 4  & -52     & 40  & 0.5 & 0.7 & 1.0, 2.4 & 106  & 57   & 18  & 2.7 \\ 
5b                      & 22.928          & -0.252   & 4.9, (10.8)    & 3    & 2  & -24     & 40  & 0.5 & 0.7 & 1.1, 1.4 & 32   & 18   & 7   & 2.0 \\ 
6                       & 23.389          & -0.187   & 9.2            & 9    & 7  & 63      & 29  & 0.6 & 0.7 & 1.6      & 84   & 34   & 13  & 3.8 \\ 
7a                      & 23.993          & 0.110    & 5.8, (9.8)     & 5    & 3  & 0       & 37  & 0.5 & 0.6 & 1.4, 2.2 & 18   & 8    & 4   & 2.3 \\ 
7b                      & 23.600          & 0.166    & 5.8, (9.8)     & 7    & 4  & 0       & 37  & 0.5 & 0.6 & 0.7, 1.7 & 38   & 19   & 9   & 2.2 \\ 
8                       & 24.152          & -0.374   & 10.3           & 13   & 11 & -68     & 114 & 1.7 & 2.0 & 2.3      & 392  & 315  & 137 & 0.6 \\ 
10                      & 25.058          & 0.165    & 6.1            & 15   & 11 & 72      & 108 & 1.5 & 1.7 & 1.2      & 690  & 341  & 165 & 2.1 \\ 
12                      & 25.995          & 0.127    & 8.8, (6.5)     & 7    & 5  & 0       & 107 & 1.6 & 1.8 & 1.8, 1.4 & 184  & 66   & 34  & 3.4 \\ 
17                      & 30.610          & 0.167    & 6.3            & 24   & 16 & -13     & 41  & 0.6 & 0.7 & 0.7      & 409  & 266  & 61  & 2.3 \\ 
19                      & 34.294          & 0.220    & 2.2            & 9    & 5  & -30     & 27  & 0.6 & 0.7 & 0.6      & 48   & 24   & 10  & 2.3 \\ 
22                      & 37.114          & -0.577   & 10.5           & 7    & 4  & -41     & 37  & 1.5 & 1.6 & 2.2      & 55   & 18   & 11  & 3.2 \\ 
23                      & 38.294          & -0.004   & 3.5, (9.9)     & 6    & 6  & 0       & 35  & 1.3 & 1.4 & 1.1, 2.4 & 48   & 24   & 13  & 1.9 \\ 
25                      & 49.318          & -0.215   & 5.7            & 7    & 6  & -38     & 155 & 0.6 & 0.8 & 0.6      & 198  & 124  & 28  & 2.6 \\ 
26                      & 283.976         & -0.868   & 4              & 19   & 10 & -79     & 61  & 0.4 & 0.5 & 0.2      & 278  & 196  & 47  & 1.7 \\ 
28a                     & 291.177         & -0.952   & 7.4            & 7    & 6  & 0       & 121 & 0.8 & 1.0 & 0.7      & 175  & 92   & 28  & 3.0 \\ 
28b                     & 292.238         & -0.666   & 7.4            & 11   & 9  & -23     & 121 & 0.8 & 1.0 & 1.0      & 333  & 193  & 47  & 3.0 \\
29\tablenotemark{a}     & 298.550         & -0.720   & 9.7            & 11   & 10 & 0       & 100 & 1.0 & 1.4 & 1.0      & 897  & 491  & 102 & 4.0 \\ 
30                      & 311.034         & 0.454    & 3.5            & 7    & 4  & 67      & 63  & 0.4 & 0.5 & 0.9      & 43   & 34   & 10  & 0.9 \\ 
31                      & 311.539         & 0.347    & 7.4            & 8    & 7  & -63     & 48  & 0.3 & 0.4 & 1.2      & 87   & 45   & 13  & 3.4 \\ 
33                      & 328.013         & -0.568   & 3.7            & 23   & 7  & -7      & 193 & 0.4 & 0.6 & 0.3      & 735  & 502  & 138 & 1.7 \\ 
34a                     & 333.514         & -0.080   & 3.4            & 6    & 5  & 0       & 604 & 0.3 & 0.7 & 0.2      & 473  & 299  & 70  & 2.5 \\ 
34b                     & 332.931         & -0.569   & 3.4            & 17   & 7  & -23     & 604 & 0.3 & 0.7 & 0.3      & 1766 & 1239 & 212 & 2.5 \\ 
36                      & 336.537         & -0.162   & 5.4, (10.2)    & 6    & 4  & -50     & 26  & 1.0 & 1.1 & 0.8, 2.0 & 36   & 12   & 6   & 4.1 \\ 
37                      & 337.243         & -0.011   & 10.9           & 10   & 6  & 9       & 495 & 1.3 & 1.8 & 1.6      & 781  & 460  & 91  & 3.5 \\ 
38                      & 337.787         & -0.382   & 3.5            & 9    & 7  & -60     & 44  & 0.4 & 0.5 & 0.9      & 91   & 50   & 16  & 2.6 \\ 
39                      & 338.361         & 0.197    & 13.3, (2.5)    & 7    & 7  & 90      & 14  & 1.8 & 1.9 & 1.8, 0.2 & 24   & 11   & 5   & 2.4 \\ 
\enddata
\tablenotetext{a}{Candidate cluster properties from \citet{rahman11a}}
\tablenotetext{c}{Where a kinematic distance ambiguity exists, we list 
the distance most consistent with candidate extinction first, and the 
less consistent distance in parentheses. }
\tablenotetext{c}{Model extinction from \citet{marshall06}}
\end{deluxetable*}

%% file: tab2-emapj.tex
\begin{turnpage}
%\begin{center}
\begin{deluxetable*}{cccccccccccc}
\tablenum{2}\label{tab:physprop}
%\tablewidth{0pt} 
%\tabletypesize{\small}
\tablecaption{Estimated Physical Properties of Candidate
  Clusters/Associations} \tablehead{
  \colhead{WMAP} & \colhead{SFC} & \colhead{D} &
  \colhead{$\overline{A_K}$} & \colhead{$N_{sub}$} &
  \colhead{Morphology} & \colhead{Latest} &
  \colhead{$M_{\textrm{min}}$} & \colhead{Log $M_{*}$\tablenotemark{b}} & \colhead{Log
    Q$_{0}$\tablenotemark{b}} & \colhead{S$_{\textrm{ff}}$ (Est)\tablenotemark{b,c}} &
  \colhead{S$_{\textrm{ff}}$ (WMAP)\tablenotemark{c}}\\ \colhead{Source} & \colhead{} &
  \colhead{(kpc)} & \colhead{(mag)} & \colhead{} &
  \colhead{Flag\tablenotemark{a}} & \colhead{Stellar Type} &
  \colhead{(M$_{\sun}$)} & \colhead{(M$_{\sun}$)} &
  \colhead{(s$^{-1}$)} & \colhead{(Jy)} & \colhead{(Jy)} }

\startdata
G24                   & 5a  & 4.9\tablenotemark{*}  & 0.6  & 49$\pm$18   & C & B9V/A0V   & 2.6  & 3.0      & [49.8]    & [144]     &      \\
                      & 5b  & 4.9\tablenotemark{*}  & 0.6  & 14$\pm$7    & C & B9V/A0V   & 2.6  & 2.4      & [49.2]    & [41]      &      \\
                      & 6   & 9.2                      & 0.65 & 50$\pm$13   & C & B4V       & 5.5  & 3.4      & 50.2      & 115       &      \\
                      & 7a  & 5.8\tablenotemark{*}  & 0.55 & 10$\pm$4    & S & B9V       & 2.7  & 2.3      & [49.1]    & [22]      &      \\
                      & 7b  & 5.8\tablenotemark{*}  & 0.55 & 19$\pm$9    & S & B9V       & 2.7  & 2.6      & [49.4]    & [42]      &      \\
                      & 8   & 10.3                     & 1.85 & 77$\pm$137  & C & O9.5V/B0V & 15   & 4.2      & 51.0      & 576       &      \\
                      & 10  & 6.1                      & 1.6  & 349$\pm$165 & C & B5V       & 5    & 4.2      & 51.0      & 1610      &      \\
                      & 12  & 8.8\tablenotemark{*}  & 1.7  & 118$\pm$34  & C & B0.5V     & 13   & 4.3      & 51.1      & 985       &      \\
\em{Total}            &     &                          &      &             &   &           &      & \em{4.8} & \em{51.6} & \em{3535} & 1377 \\
\hline
G30                   & 17  & 6.3                      & 0.65 & 143$\pm$61  & C & B9V       & 2.7  & 3.5      & 50.3      & 267       & 1585 \\
\hline
G34                   & 19  & 2.2                      & 0.65 & 24$\pm$10   & C & A8V       & 2    & 2.5      & [49.3]    & [243]     & 285  \\
\hline
G37                   & 22  & 10.5                     & 1.55 & 37$\pm$11   & S & B0V       & 14   & 3.9      & 50.7      & 241       &      \\
                      & 23  & 3.5\tablenotemark{*}  & 1.35 & 24$\pm$13   & C & B9V/A0V   & 2.6  & 2.7      & [49.5]    & [138]     &      \\
\em{Total}            &     &                          &      &             &   &           &      & \em{3.9} & \em{50.7} & \em{380}  & 244  \\
\hline
G49                   & 25  & 5.7                      & 0.7  & 74$\pm$28   & C & B9V       & 2.7  & 3.2      & 50.0      & 169       & 458  \\
\hline
G283                  & 26  & 4                        & 0.45 & 82$\pm$47   & S & A0V       & 2.5  & 3.2      & 50.0      & 341       & 848  \\
\hline
G291                  & 28a & 7.4                      & 0.9  & 83$\pm$28   & S & B8V/B9V   & 3    & 3.3      & 50.1      & 130       &      \\
                      & 28b & 7.4                      & 0.9  & 140$\pm$47  & S & B8V/B9V   & 3    & 3.5      & 50.3      & 218       &      \\
\em{Total}            &     &                          &      &             &   &           &      & \em{3.7} & \em{50.5} & \em{348}  & 688  \\
\hline
G298\tablenotemark{b} & 29  & 9.7                      & 1.1  & 406$\pm$102 & C & O9.5V     & 16.5 & 5.0      & 51.8      & 307       & 313  \\
\hline
G311                  & 30  & 3.5                      & 0.45 & 9$\pm$10    & C & A1V       & 2.3  & 2.2      & [49.0]    & [44]      &      \\
                      & 31  & 7.4                      & 0.35 & 42$\pm$13   & S & B9V       & 2.7  & 2.9      & [49.7]    & [57]      &      \\
\em{Total}            &     &                          &      &             &   &           &      & \em{3.0} & \em{49.8} & \em{102}  & 766  \\
\hline
G327                  & 33  & 3.7                      & 0.5  & 233$\pm$138 & S & A1V       & 2.3  & 3.6      & 50.4      & 1014      & 943  \\
\hline
G332                  & 34a & 3.4                      & 0.5  & 174$\pm$70  & C & A1V       & 2.3  & 3.5      & 50.3      & 899       &      \\
                      & 34b & 3.4                      & 0.5  & 527$\pm$212 & C & A1V       & 2.3  & 3.9      & 50.7      & 2714      &      \\
\em{Total}            &     &                          &      &             &   &           &      & \em{4.1} & \em{50.9} & \em{3613} & 1787 \\
\hline
G337                  & 36  & 5.4\tablenotemark{*}  & 1.05 & 24$\pm$6    & C & B9V       & 2.7  & 2.7      & [49.5]    & 62        &      \\
                      & 37  & 10.9                     & 1.55 & 321$\pm$91  & C & B0V       & 14   & 4.8      & 51.6      & 1944      &      \\
                      & 38  & 3.5                      & 0.45 & 41$\pm$16   & S & A1V       & 2.3  & 2.8      & [49.6]    & [201]     &      \\
                      & 39  & 13.3\tablenotemark{*} & 1.85 & 13$\pm$5    & C & O8.5V     & 20   & 3.6      & 50.4      & 86        &      \\
\em{Total}            &     &                          &      &             &   &           &      & \em{4.8} & \em{51.6} & \em{2293} & 2239 \\
\enddata

\tablecomments{The Q$_{0}$ and S$_{ff}$ values enclosed in brackets
  are likely incorrect due to poor sampling of the IMF (described in
  \S\ref{sect:phys}.)}

\tablenotetext{a}{C refers to a candidate that appears as a central
  cluster, S refers to a candidate that appears embedded in a shell
  structure.}  \tablenotetext{b}{Candidate cluster properties from
  \citet{rahman11a}.}\tablenotetext{c}{The error on these parameters
  scale linearly with the error on $N_{sub}$ } \tablenotetext{d}{The
  flux measurements are given at 90 GHz.} \tablenotetext{*}{The
  kinematic distance that is most consistent with the expected average
  distance-to-extinction ratio is used.}
\end{deluxetable*}
%\end{center}
\end{turnpage}

%% file: tab3-emapj.tex
\begin{deluxetable*}{ccccc}
%\tablewidth{0pt}
\tablenum{3}\label{tab:census} 
\tabletypesize{\scriptsize} \tablecaption{Census of
  Massive ($\ga 10^4$ M$_{\sun}$), Young ($\la 4$ Myr) Clusters \& OB
  Associations in the Galaxy}
\tablehead{ \colhead{Galactic Region} & \colhead{Mass (log M$_{\sun}$)} &
  \colhead{Distance (kpc)} & \colhead{Name} & \colhead{Reference}}
\startdata
Galactic Centre & 4.1                       & 8.5                      & Central Cluster & (5)  \\
                & 4.0                       & 8.5                      & Arches          & (3)  \\
                & 4.0                       & 8.5                      & Quintuplet      & (3)  \\
\hline
G24             & $\ge$4.2\tablenotemark{*} & 10.3                     & SFC 8           & (1)  \\
                & $\ge$4.2\tablenotemark{*} & 6.1                      & SFC 10          & (1)  \\
                & $\ge$4.3\tablenotemark{*} & 8.8\tablenotemark{\#}     & SFC 12          & (1)  \\
\hline
G37             & $\ge$3.9\tablenotemark{*} & 10.5                     & SFC 22          & (1)  \\
\hline
G43             & $\sim$4                   & 11.4                     & W49A            & (2)  \\
\hline
G80             & 4.4                       & 1.5                      & Cygnus OB2      & (8) \\
\hline
G283            & $>$4.0                    & 2.8                      & RCW49, Westerlund2 & (9) \\
\hline
G287            & $>$4.3                    & 2.3                      & Carina          & (7) \\
\hline
G291            & 4.2                       & 6.0                      & NGC 3603        & (4)  \\
\hline
G298            & 5.0                       & 9.7                      & Dragonfish      & (6)  \\
\hline
G332            & $\ge$4.1\tablenotemark{*} & 3.2                      & SFC 34a,b       & (1)  \\
\hline
G337            & $\ge$4.8\tablenotemark{*} & 10.9                     & SFC 37          & (1)  \\
\enddata
\tablenotetext{*}{Candidate Cluster/Association from this work.}
\tablenotetext{\#}{Unresolved kinematic distance ambiguity. The
  distance indicated is the most consistent with the model extinction
  to the candidate.} 
\tablerefs{
(1) {\em This work};
(2) \citet{conti02};
(3) \citet{figer99};
(4) \citet{harayama08}; 
(5) \citet{paumard06};
(6) \citet{rahman11a};
(7) \citet{smith07};
(8) \citet{wright10};
(9) \citet{ascenso07};
}
\tablecomments{\citet{wolff07} use an IMF-sensitive extrapolation over
  a small mass range to determine the masses of IC 1805 and NGC
  6611. The extrapolated mass of NGC 6611 is $10^{4.4}$ M$_{\sun}$, in
  contrast to more recent work by \citet{bonatto06} that revise the
  minimum cluster mass down to $10^{3.2}$ M$_{\sun}$. In both of these
  cases, the free-free luminosities of the regions from
  \citet{murray10} are inconsistent with young clusters with M$ >
  10^4$ M$_{\sun}$, indicating smaller masses. Consequently, we
  exclude both of these regions from this census.}

\end{deluxetable*}

%% file: tab4-emapj.tex
\begin{deluxetable*}{ccc}

%\tablewidth{0pt} \tabletypesize{\scriptsize} 
\tablenum{4}\label{tab:missing}
\tablecaption{Giant \hii
  Regions Lacking Known Young Massive Clusters}

\tablehead{ \colhead{Galactic Region} & \colhead{Distance Components
    (kpc)} & \colhead{Names}}

\startdata 
G3                       & 14.3, 14.6 & AMWW34, AMWW35 \\ 
G6\tablenotemark{*}   & 2.8        & M8             \\ 
G8                       & 13.5       & \nodata        \\ 
G10\tablenotemark{*}  & 4.5, 15    & W31            \\ 
G15\tablenotemark{*}  & 2.4        & M17            \\
G20                      & 11.8       & \nodata        \\ 
G30\tablenotemark{\#} & 6.3        & W43, SFC 17    \\ 
G49\tablenotemark{\#} & 5.7        & W51, SFC 25    \\ 
G70                      & 8.6        & W58A           \\ 
G133\tablenotemark{*} & 4.2        & W3             \\ 
G274                     & 6.4        & RCW 42         \\ 
G289                     & 7.9        & \nodata        \\ 
G305\tablenotemark{*} & 3.5        & \nodata        \\ 
G320                     & 11.5, 12.6 & \nodata        \\ 
G327\tablenotemark{\#}& 3.6        & RCW97, SFC 33  \\
G347\tablenotemark{*} & 7.9        & [DBS2003] 179  \\ 
G351                     & 13.7       & \nodata        \\ 
\enddata

\tablenotetext{*}{Regions with known powering clusters, but
  significantly less massive than $10^4$ M$_{\odot}$ }
\tablenotetext{\#}{Regions with candidate powering clusters, but
  with minimum stellar masses less than $10^4$ M$_{\odot}$ }
\tablecomments{All regions and distance components are from
  \citet{conti04}. }

\end{deluxetable*}